%% file: HARX_SM.tex
\documentclass[twocolumn]{autart}
\usepackage{etex}

\usepackage{IEEEtrantools}
\input{packages}
\input{preNiklas}

\input{system_connections_tikz}

\begin{document}

\begin{frontmatter}

\title{Optimal model order reduction with the Steiglitz-McBride method for open loop data\thanksref{footnoteinfo}} % Title, preferably not more than 10 words.

\thanks[footnoteinfo]{This work was  supported by the Swedish Research Council under contract 2015-05285 and by the European Research Council under the advanced grant LEARN, contract 267381.
The material in this paper was not presented at any conference.
}

\author[kth]{Niklas Everitt}\ead{everitt@kth.se},    % Add the
\author[kth]{Miguel Galrinho}\ead{galrinho@kth.se},               % e-mail address
\author[kth]{H{\aa}kan Hjalmarsson}\ead{hjalmars@kth.se}  % (ead) as shown

\address[kth]{ACCESS Linnaeus Center, School of Electrical Engineering, KTH - Royal Institute of Technology, Sweden}


\begin{abstract}                          % Abstract of not more than 200 words.
%Estimating Box-Jenkins models with the prediction error method (PEM) requires solving a non-convex optimization problem.
In system identification, it is often difficult to find a physical intuition to choose a noise model structure.
The importance of this choice is that, for the prediction error method (PEM) to provide asymptotically efficient estimates, the model orders must be chosen according to the true system.
However, if only the plant estimates are of interest and the experiment is performed in open loop, the noise model may be over-parameterized without affecting the asymptotic properties of the plant.
The limitation is that, as PEM suffers in general from non-convexity, estimating an unnecessarily large number of parameters will increase the chances of getting trapped in local minima.
To avoid this, a high order ARX model can first be estimated by least squares, providing non-parametric estimates of the plant and noise model.
Then, model order reduction can be used to obtain a parametric model of the plant only.
We review existing methods to perform this, pointing out limitations and connections between them.
Then, we propose a method that connects favorable properties from the previously reviewed approaches.
We show that the proposed method provides asymptotically efficient estimates of the plant with open loop data.
Finally, we perform a simulation study, which suggests that the proposed method is competitive with PEM and other similar methods.
\end{abstract}

\end{frontmatter}

\section{Introduction}

The prediction error method (PEM) is a well-know approach for estimation of parametric models~\cite{ljung99}. If the model orders are chosen correctly, a quadratic cost function provides asymptotically efficient estimates when the noise is Gaussian. The drawback is that, in general, PEM requires solving a non-convex optimization problem, which can converge to minima that are only local.  
Alternative methods, such as subspace~\cite{subspacebook} or instrumental variable~\cite{ivbook_soderstrom_stoica} methods, do not suffer from non-convexity, being useful to provide initialization points for PEM.
%However, they often do not provide optimal estimates.

Other methods first estimate a high order (non-parametric) model. 
In general, this is an ARX model, for which the global minimum of the prediction error cost function can be found by least squares.
Because it is high order, this estimate will have high variance.
However, it can be reduced to a parametric model description of low order.
If the model reduction step is performed according to an exact maximum likelihood (ML) criterion, the low order estimates are asymptotically efficient~\cite{wahlberg89}.
This approach still requires, in general, solving a non-convex optimization problem.

Another possibility to perform model order reduction from a high order non-parametric model is with the weighted null-space fitting (WNSF) method~\cite{Galrinho:14}.
Although it can be motivated by an exact ML criterion, this criterion is not minimized explicitly.
Rather, it is interpreted as a weighted least squares problem by fixing the parameters in the weighting.
%Then, iterative weighted least squares is applied.

One problem with estimation of parametric models is the choice of model orders.
If this choice can sometimes be based on physical intuition for the plant, the noise model order is usually a more abstract concept.
This has been observed in~\cite{freqdom_nonparH}, where a frequency domain method is proposed to estimate a parametric model of the plant and a non-parametric noise model.
Because this approach does not require a noise model order selection, it can be seen as more user friendly.

If the data are obtained in open loop, the asymptotic properties of the plant and noise model estimates obtained with PEM are uncorrelated, if the two transfer functions are independently parametrized~\cite{ljung99}.
Therefore, when a parametric noise model estimate is not of interest, asymptotically efficient estimates of the plant can be obtained as long as the noise model order is chosen high enough for the system to be in the model set.
The limitation of choosing the noise model order arbitrarily large with PEM is that, as more parameters are estimated, the complexity of the problem increases, and it is more difficult to find the global minimum.
%For WNSF, the complexity of the model reduction step also increases when the order of the noise model is chosen to be high.

However, if a non-parametric ARX model is estimated, there are no issues with local minima, while the order is arbitrarily large.
Then, for the model reduction step, an approximate asymptotic ML criterion allows separating the estimation of the plant and noise model~\cite{wahlberg89}.
This allows obtaining asymptotically efficient estimates of the plant in open loop without the high order structure of the noise model affecting the difficulty of the problem.
Nevertheless, the model reduction step still requires solving a non-convex optimization problem.
The ASYM method~\cite{zhu_book} is based on this approach.

Another approach that does not require a parametric noise model is the BJSM method~\cite{zhu_hakan}.
This method uses a non-parametric ARX model estimate to pre-filter the input and output data, creating a pre-filtered data set for which the output noise is approximately white.
Then, a noise model is no longer required when estimating the plant based on the pre-filtered data set. 
Instead of explicitly minimizing a non-convex function, BJSM applies the Steiglitz-McBride to the pre-filtered data set.
In~\cite{zhu_hakan}, it is shown that this procedure is asymptotically efficient in open loop.
However, there are two limitations.
First, even if the true noise model is known exactly, a non-parametric estimate is still required to achieve efficiency.
Second, although the method does not apply local non-linear optimization techniques, the number of Steiglitz-McBride iterations needs to tend to infinity to obtain a consistent estimate.

Our contributions are the following.
First, we make a connection between ASYM and BJSM, and propose a method---termed Model Order Reduction Steiglitz-McBride (MORSM)---connecting ideas from both.
Second, we show that MORSM is asymptotically efficient in open loop with one iteration.
Third, we perform a simulation study, where we observe that MORSM has better finite sample convergence properties than BJSM, and that it is a viable alternative to PEM.

\section{Preliminaries}
\label{sec:preliminaries}
\input{preliminaries}

\section{The Prediction Error Method}
\label{sec:pem}
\input{pem}

\section{Model Reduction}
\label{sec:using-ARX}
\input{using-ARX}

% \section{Maximum Likelihood}
% \label{sec:ml}
% \input{ml}

\section{Model Order Reduction Steiglitz-McBride}
\label{sec:method}
\input{method}

\section{Asymptotic Properties}
\label{sec:properties}
\input{properties}

\section{Simulations}
\label{sec:sim}
\input{simulation}

\section{Conclusion}
In this paper, we propose a least squares method for estimation of models with a plant parameterized by a rational transfer function and a non-parametric noise model. 
We show that the method provides consistent and asymptotically efficient estimates of the plant if data are obtained in open loop.

Essentially, the method performs model order reduction based on an asymptotic ML criterion using the Steiglitz-McBride method.
We thus name it Model Order Reduction Steiglitz-McBride (MORSM).
The method uses ideas from the ASYM and BJSM methods. 
However, unlike ASYM, we avoid a non-convex optimization procedure by applying Steiglitz-McBride; unlike BJSM, we propose a procedure that only requires one iteration to provide asymptotically efficient estimates.

Finally, we perform two simulation studies to analyze the performance of the method, from which the following are observed. 
First, MORSM is asymptotically efficient in one iteration, while BJSM is not. 
Second, even when extra iterations are required for convergence with finite sample sizes, MORSM still converges in less iterations than BJSM. 
Third, MORSM may be a viable alternative to PEM, specially when PEM has difficulty in finding the global minimum.

Future work will include application of MORSM for closed loop and for estimation of systems embedded in networks.

\appendix
\input{appendices}

\bibliographystyle{plain}        % Include this if you use bibtex
\bibliographystyle{IEEEbib}
\bibliography{HARX_SM_bib}

\end{document}

%% file: packages.tex
\usepackage{cite}
\usepackage{booktabs} % for much better looking tables
\usepackage{array} % for better arrays (eg matrices) in maths
\usepackage{paralist} % very flexible & customisable lists (eg. enumerate/itemize, etc.)
\usepackage{verbatim} % adds environment for commenting out blocks of text & for better verbatim
\usepackage{mathtools}
\usepackage{mathrsfs}
\usepackage{amsfonts}
\usepackage{nicefrac}
\usepackage{amsmath}
\usepackage{amssymb}
\usepackage{bbm}     % Used for the indicator function

%% file: preNiklas.tex
\newcommand{\transpose}{^\top}
\newcommand{\nul}{\circ}

\newcommand{\ie}{i.e.,~}
\newcommand{\eg}{e.g.,~}
\renewcommand{\j}{\mathrm{j}}

\newcommand{\eiw}{\e^{\j\omega}}

\newcommand{\proj[1]}{\mathbf{P}_{#1}\!}
\newcommand{\AsN}{\operatorname{AsN}}
\renewcommand{\Psi}{\varPsi}
\renewcommand{\Gamma}{\varGamma}
\renewcommand{\Lambda}{\varLambda}
\renewcommand{\Phi}{\varPhi}
\renewcommand{\Omega}{\varOmega}
\renewcommand{\Sigma}{\varSigma}
\renewcommand{\Theta}{\varTheta}
\renewcommand{\Pi}{\varPi}
\renewcommand{\Upsilon}{\varUpsilon}

\newcommand{\beq}{\begin{equation}}
\newcommand{\eeq}{\end{equation}}
\newcommand{\beqn}{\begin{equation*}}
\newcommand{\eeqn}{\end{equation*}}
\newcommand{\beqarr}[1]{\begin{IEEEeqnarray}{#1}}
\newcommand{\eeqarr}{\end{IEEEeqnarray}}
\newcommand{\beqarrn}[1]{\begin{IEEEeqnarray*}{#1}}
\newcommand{\eeqarrn}{\end{IEEEeqnarray*}}
\newcommand{\bbmat}{\begin{bmatrix}}
\newcommand{\ebmat}{\end{bmatrix}}

\newcommand{\norm}[1]{\left \lVert #1 \right \rVert}

\newcommand{\wint}[1]{\int_{-\pi}^{\pi} \!\!\!  #1 \: \mathrm{d} \omega}
\newcommand{\expect}[1]{\mathrm{E}\left [ #1 \right]}

\newcommand{\inp}[1]{\left \langle #1 \right\rangle}
\newcommand{\defeq}{\vcentcolon=} %\mathrel{\mathop:}

\newcommand{\abs}[1]{\left\vert #1 \right\vert}

\newcommand{\bnul}{\begin{enumerate}[a)]}
\newcommand{\enul}{\end{enumerate}}

\newtheorem{theorem}{Theorem}[section]

\newtheorem{lemma}{Lemma}[section]
\newtheorem{assumption}{Assumption}[section]

\newcounter{counter}

\newcommand{\Fc}{{\mathcal F}}

\newcommand{\Lc}{{\mathcal L}}

\newcommand{\Oc}{{\mathcal O}}

\newcommand{\Sc}{{\mathcal S}}

\newcommand{\Rb}{{\mathbb R}}

%% file: system_connections_tikz.tex
\usepackage{tikz} % Tikz! rita rita
\usepackage{pgfplots}
\pgfplotsset{compat=newest}
\usetikzlibrary{calc,intersections,through,backgrounds,shapes.misc,positioning,chains,scopes}
\tikzset{subsystem/.style={
rectangle,
minimum size=6mm,
thick,
draw=black,%red!50!black!50, % 50% red and 50% black,
top color=white, % a shading that is white at the top...
bottom color=white, %red!50!black!20, % and something else at the bottom
font=\itshape,
text height=1.5ex,
text depth=0.25ex
}}

\tikzset{operator/.style={
circle,minimum size=15pt,%rounded corners=6pt,
inner sep=0pt,
thin,draw=black, %!50,
top color=white,bottom color=white,
text height=1.75ex,
text depth=0.25ex,
font=\itshape}}

\tikzset{signal/.style={
circle,minimum size=9pt,%rounded corners=6pt,
inner sep=0pt,
thin,draw=black, %!50,
top color=white,bottom color=white,
text height=1.5ex,
text depth=0.25ex,
font=\itshape}}

\tikzset{point/.style={circle,inner sep=0pt,minimum size=0pt,fill=black}}
\tikzset{skip loop/.style={to path={-- ++(0,#1) -| (\tikztotarget)}}}

%% file: preliminaries.tex
\begin{assumption}[True system]
\label{ass:true_system}
The system has scalar input $u_t$, scalar output $y_t$ and is subject to scalar noise $e_t$. The relationship between these signals is given by
\begin{IEEEeqnarray}{rCl}
\label{eq:true_BJ}
y_t &= G^\nul(q)u_t + H^\nul(q) e_t,
\label{eq:truesys}
\end{IEEEeqnarray}
where $G^\nul(q)$ and $H^\nul(q)$ are rational functions in the time shift operator $q^{-1}$ ($q^{-1}x_t := x_{t-1}$) according to
\begin{IEEEeqnarray*}{rCcCl}
G^\nul(q) &=& \frac{L^\nul(q)}{F^\nul(q)} &=& \frac{l_1^\nul q^{-1} + \dotsb + l_{m_l}^\nul q^{-m_l^\nul} }{1+ f_1^\nul q^{-1} + \dotsb + f_{m_f}^\nul q^{-m_f^\nul}},
\\
H^\nul(q) &=& \frac{C^\nul(q)}{D^\nul(q)} &=& \frac{1 + c_1^\nul q^{-1} + \dotsb + c_{m_c}^\nul q^{-m_c^\nul} }{1+ d_1^\nul q^{-1} + \dotsb + d_{m_d}^\nul q^{-m_d^\nul}}.
\end{IEEEeqnarray*}
The transfer functions $G^\nul$, $H^\nul$, and $1/H^\nul$ are assumed to be stable. 
The polynomials $L^\nul$ and $F^\nul$---as well as $C^\nul$ and $D^\nul$---do not share common factors.

\end{assumption}

Let the input sequence $\{u_t\}$ be a realization of a stochastic process generated by a random sequence $\{w_t\}$.
Also, let $\Fc_{t-1}$ be the $\sigma$-algebra generated by ${\{e_s, w_s, s\le t-1\}}$. 
Then, the following assumption applies for the input signal.

\begin{assumption}[Input]
\label{ass:input}
The sequence $\{u_t\}$ is defined by
\begin{IEEEeqnarray*}{rCl}
u_t &=& F_u(q) w_t,
\end{IEEEeqnarray*}
where $F_u(q)$ is a stable and inversely stable finite dimensional filter, where $\{w_t\}$ is independent of $\{e_t\}$, satisfying
\begin{IEEEeqnarray*}{rClrClrCl}
\expect{w_t|\Fc_{t-1}} &=& 0, \quad & \expect{w_t^2|\Fc_{t-1}} &=& \sigma_\nul^2,
\quad & |w_t| &\le & C, \forall t
\end{IEEEeqnarray*}
for some finite positive finite constant $C$.
\end{assumption}

Assumption~\ref{ass:input} implies that the system is operating in open loop.
Also, $F_u$ can be interpreted as the stable minimum phase spectral factor of the input spectrum.

For the noise, the following assumption applies.

\begin{assumption}[Noise]
\label{ass:noise}
 $\{e_t\}$ is a stochastic process that satisfies
\begin{IEEEeqnarray*}{rClrClrCl}
\expect{e_t|\Fc_{t-1}} &=& 0, \quad & \expect{e_t^2|\Fc_{t-1}} &=& \sigma_\nul^2,
\quad & |e_t|^{10} &\le & C, \forall t
\end{IEEEeqnarray*}
for some positive finite constant $C$.
\end{assumption}

% \begin{definition}[$f_N$-quasi-stationarity]
% Let $f_N$ be a decreasing sequence of positive scalars, such that $f_N \to 0$ as $N \to \infty$. Define
% \begin{IEEEeqnarray*}{rCl}
% R^N_{vv}(t) &=& \left \{ \begin{array}{ll}
% \frac{1}{N}\sum_{t = \tau+1}^N v_tv_{t_\tau}\transpose, & 0 \le \tau < N \\
% \frac{1}{N}\sum_{t = 1}^{N+\tau} v_tv_{t_\tau}\transpose, & -N < \tau \le 0\\
% 0 & \mathrm{otherwise.}
% \end{array}
% \right.
% \end{IEEEeqnarray*}
% The vector sequence $\{v_t\}$ is said to be $f_n$-quasi-stationary if
% \begin{enumerate}[i)]
% \item There exists $\{R_{vv}(\tau)\}$ such that \\ $\sup_{|\tau| \le N} \norm{R^N_{vv}(\tau)- R_{vv}(\tau)}_2 \le C_1 f_N$
% \item $\frac{1}{N}\sum_{t = -N}^N \norm{v_t}_2^2  \le C_2$
% \end{enumerate}
% for $N$ large enough, where $C_1$ and $C_2$ are finite constants.
% \end{definition}

% \begin{definition}[$f_N$-stability]
% A filter $G(q) = \sum_{k = 0}^\infty g_k q^{-k}$ is said to be $f_N$ stable if
% \begin{IEEEeqnarray*}{rCl}
% \sum_{k = 0}^\infty \norm{g_k}_2 /f_k < \infty
% \end{IEEEeqnarray*}
% \end{definition}

%% file: pem.tex
The idea of the prediction error method (PEM) is to minimize a cost function of the prediction errors.
In this section, we discuss how PEM can be used to estimate a model of the system~\eqref{ass:true_system}.
First, we consider a Box-Jenkins (BJ) model, and then a high order ARX model.

\subsection{Box-Jenkins model}

In a Box-Jenkins model, $G(q)$ and $H(q)$ are rational transfer functions parameterized independently, according to
\begin{IEEEeqnarray}{rCl}
\label{eq:model_BJ}
y_t &= G(q,\theta)u_t + H(q,\alpha) e_t,
\end{IEEEeqnarray}
where 
\begin{IEEEeqnarray*}{rCcCl}
G(q,\theta) &=& \frac{L(q,\theta)}{F(q,\theta)} &=& \frac{l_1 q^{-1} + \dotsb + l_{m_l} q^{-m_l} }{1+ f_1 q^{-1} + \dotsb + f_{m_f} q^{-m_f}},
\\
H(q,\alpha) &=& \frac{C(q,\alpha)}{D(q,\alpha)} &=& \frac{1 + c_1 q^{-1} + \dotsb + c_{m_c} q^{-m_c} }{1+ d_1 q^{-1} + \dotsb + d_{m_d} q^{-m_d}},
\end{IEEEeqnarray*}
and
\begin{IEEEeqnarray}{rCl}
\theta & = &
\begin{bmatrix}
f_1 & \dotsc & f_{m_f} & l_1 & \dotsc & l_{m_l} 
\end{bmatrix}\transpose , \\
\alpha & = & 
\begin{bmatrix}
c_1 & \dotsc & c_{m_c} & d_1 & \dotsc & d_{m_d} 
\end{bmatrix}\transpose .
\end{IEEEeqnarray}
We assume that $H^\nul(q)$ is in the model set defined by $H(q,\alpha)$ (i.e., $m_c\geq m_c^\nul$ and $m_d\geq m_d^\nul$).
Moreover, the order of the polynomials of $G^\nul(q)$ are assumed to be known (i.e., $m_f=m_f^\nul$ and $m_l=m_l^\nul$).
For simplicity of notation only, we also assume that $m:=m_f=m_l$.

The one step ahead prediction errors of the BJ model~\eqref{eq:model_BJ} are given by
\begin{IEEEeqnarray*}{rCl}
\varepsilon_t(\theta,\alpha) &=& \frac{D(q,\alpha)}{C(q,\alpha)} \left[ y_t - \frac{L(q,\theta)}{F(q,\theta)}u_t \right] .
\end{IEEEeqnarray*}
The parameter estimates using PEM with a quadratic cost function are determined by minimizing the loss function
\begin{IEEEeqnarray}{rCl}
V_N(\theta,\alpha) &=& \frac{1}{N}\sum_{t=1}^N \varepsilon_t^2(\theta,\alpha),
\label{eq:lossfunc}
\end{IEEEeqnarray}
where $N$ is the number of data samples.
We denote by $\hat{\theta}_N^\mathrm{PEM}$ the estimate of $\theta$ obtained by minimizing~\eqref{eq:lossfunc}.
Moreover, $\theta_\nul$ corresponds to the vector $\theta$ evaluated at the coefficients of $F^\nul(q)$ and $L^\nul(q)$.

Since the system operates in open loop (Assumption~\ref{ass:input}), it is well known that, when PEM is applied to the model~\eqref{eq:model_BJ}, the asymptotic covariance matrix of the parameter estimate $\hat{\theta}_N^\mathrm{PEM}$ is given by~\cite{ljung99}
\begin{IEEEeqnarray*}{rCl}
\lim_{N \to \infty} N \expect{ ( \hat{\theta}_N^\mathrm{PEM} - \theta_\nul) ( \hat{\theta}_N^\mathrm{PEM} - \theta_\nul)\transpose } &=& \sigma_\nul^2 M_{\text{CR}}^{-1} ,
\label{eq:PEMcov}
\end{IEEEeqnarray*}
where (we omit the argument of the transfer functions for brevity)
\begin{IEEEeqnarray*}{rCl}
 M_{\text{CR}} 
%  &=& \frac{1}{\sigma_\nul^2}
%  \expect{ \begin{bmatrix}
%  - \frac{G^\nul}{F^\nul H^\nul} \alpha_m
%  \\
%  \frac{1}{F^\nul H^\nul} \alpha_m
%  \end{bmatrix}
%  u_t
%  \left( \begin{bmatrix}
%  - \frac{G^\nul}{F^\nul H^\nul} \alpha_m
%  \\
%  \frac{1}{F^\nul H^\nul} \alpha_m
%  \end{bmatrix}
%  u_t \right)^{\!\!\! \mathrm{T} }
% \:
%  }
%  \\
 &=&
 \frac{1}{2 \pi \sigma_\nul^2}
 \wint{ \: \begin{bmatrix}
 - \frac{G^\nul}{F^\nul H^\nul} \Gamma_m
 \\
 \frac{1}{F^\nul H^\nul} \Gamma_m
 \end{bmatrix}
  \begin{bmatrix}
 - \frac{G^\nul}{F^\nul H^\nul} \Gamma_m
 \\
 \frac{1}{F^\nul H^\nul} \Gamma_m
 \end{bmatrix}
 ^{\! * } \!
 \Phi_u},
\end{IEEEeqnarray*}
with $\Gamma_m(q) = \begin{bmatrix}
q^{-1} & \dotsc & q^{-m}
\end{bmatrix} \transpose$ and $\Phi_u$ the spectrum of the input $\{u_t\}$. 
%Thus, for operation in open loop, the estimate of $\theta$ is asymptotically independent of the parameters $\alpha$ in the noise model $H(q,\alpha)$. 

When $\{e_t\}$ is Gaussian, PEM with a quadratic cost function is asymptotically efficient, meaning that $M_{\text{CR}}^{-1}$ corresponds to the Cram{\'e}r-Rao lower bound---the smallest possible asymptotic covariance matrix for a consistent estimator\cite{ljung99}.
Again, we recall that only the orders of $G^\nul(q)$ need to be chosen correctly to achieve efficiency, while $H(q,\alpha)$ only needs to include $H^\nul(q)$.
Thus, if only a model for $G^\nul(q)$ is of interest, and the order of $H^\nul(q)$ is unknown, $m_c$ and $m_d$ can be let grow to infinity (guaranteeing that $H^\nul(q)$ is in the model set) without asymptotically affecting the estimate of $\theta$.

An important remark is that minimizing the loss function \eqref{eq:lossfunc} is a non-convex optimization problem.
Therefore, a good initialization point is required to converge to the global minimum. 
For Box-Jenkins models, an initialization point that is sufficiently close to the global minimum is particularly challenging to obtain. 
Moreover, the problem becomes yet more challenging if we want to let the order of the noise model $H(q,\alpha)$ be arbitrarily large, as PEM will have increased problems with local minima.

\subsection{High order ARX model}

To circumvent the limitations of solving a non-convex optimization problem, we consider the following approach. 
Note that the system~\eqref{eq:truesys} can be represented as
\begin{IEEEeqnarray}{rCl}
\label{eq:true-ARX-model}
A^\nul(q) y_t &=& B^\nul(q) u_t + e_t ,
\end{IEEEeqnarray}
where
\begin{IEEEeqnarray*}{rCcCl}
A^\nul(q) &:=& \frac{1}{H^\nul(q)} &=:& 1 + \sum_{k=1}^\infty a^\nul_k q^{-k} ,
\\
B^\nul(q) &:=& \frac{G^\nul(q)}{H^\nul(q)} &=:& \sum_{k=1}^\infty b^\nul_k q^{-k}
\end{IEEEeqnarray*}
are stable transfer functions (by Assumption~\ref{ass:true_system}).

Consider also the ARX model
\begin{IEEEeqnarray*}{rCcCl}
A(q,\eta^n) y_t &=& B(q,\eta^n)u_t + e_t ,
\label{eq:ARXmodel}
\end{IEEEeqnarray*}
where
\begin{IEEEeqnarray}{rClrCl}
\label{eq:A-B-def}
A(q,\eta^n) &=&  1 + \sum_{k=1}^n a_k q^{-k}
, \quad &
B(q,\eta^n) &=&  \sum_{k=1}^n b_k q^{-k},
\end{IEEEeqnarray}
and
\begin{equation*}
\eta^n = \begin{bmatrix}
a_1 & \dotsc & a_n & b_1 & \dotsc & b_n
\end{bmatrix}\transpose .
\end{equation*}
Here, we assumed, without loss of generality, that $A(q)$ and $B(q)$ are both modeled with $n$ coefficients. 
Note that~\eqref{eq:true-ARX-model} is not in the model set defined by~\eqref{eq:ARXmodel} due to the truncation by $n$ coefficients.
Nevertheless, the stability assumption on $G^\nul(q)$ and $1/H^\nul(q)$ implies that $\{a_k^\nul\}$ and $\{b_k^\nul\}$ are sequences converging to zero.
Thus, if $n$ is chosen large enough, \eqref{eq:ARXmodel} can model~\eqref{eq:true-ARX-model} with good accuracy. 

An advantage of ARX models is that they are linear in the model parameters.
In particular, the PEM estimate of $\eta^n$ is obtained by minimizing the cost function
\begin{IEEEeqnarray}{rCl}
V_N(\eta^n) &=& \frac{1}{N}\sum_{t=1}^N \left[ A(q,\eta^n) y_t - B(q,\eta^n) u_t \right]^2,
\end{IEEEeqnarray}
which can be done by linear least squares.
Thus, it can be solved as follows.
Write~\eqref{eq:ARXmodel} as 
\begin{IEEEeqnarray*}{rCcCl}
y_t = (\varphi_t^n)\transpose \eta^n + e_t ,
\end{IEEEeqnarray*}
where 
\begin{IEEEeqnarray}{rCl}
\label{eq:varphi-def}
\varphi_t^n &=& \begin{bmatrix}
-y_{t-1} & \dotsc & -y_{t-n} & u_{t-1} \dotsc & u_{t-n}
\end{bmatrix} \transpose \! \! .
\end{IEEEeqnarray}
Then, the least squares estimate of $\eta^n$ is given by
\begin{IEEEeqnarray}{rCl}
\label{eq:eta_N-def}
\hat{\eta}_N^{n,ls} &\defeq & [R^n_N]^{-1} r^n_N ,
\end{IEEEeqnarray}
where
\begin{IEEEeqnarray*}{rCl}
R^n_N &=& \frac{1}{N} \sum_{t = 1}^N \varphi_t^n ( \varphi_t^n)\transpose, \quad
r^n_N = \frac{1}{N} \sum_{t = 1}^N \varphi_t^n y_t .
\label{eq:RnN}
\end{IEEEeqnarray*}

In the analysis, we will use the slightly modified estimate
\begin{IEEEeqnarray}{rCl}
\hat{\eta}_N^{n} 
 &\defeq & [R^n_{N,\mathrm{reg}}]^{-1} r^n_N ,
 \label{eq:eta_reg_N-def}
\end{IEEEeqnarray}
where
\begin{IEEEeqnarray*}{rCl}
R^n_{N,\mathrm{reg}} &=&
\left\{ \begin{array}{ll}
R^n_N & \mathrm{if} \norm{[R^n_N]^{-1}}_2 <  2/ \delta \\
R^n_N + \frac{\delta}{2}I_{2n} & \mathrm{otherwise}
\end{array}
\right. ,
\end{IEEEeqnarray*}
for some small $\delta > 0$. The reason is that $\hat{\eta}_N^n$ is easier to analyze statistically, while the first and second order statistical properties of $\hat{\eta}_N^{n,ls}$ and $\hat{\eta}_N^n$ are asymptotically identical~\cite{Ljung&Wahlberg:92a}.

It follows from Assumption~\ref{ass:input} and Assumption~\ref{ass:noise} (see~\cite{Ljung&Wahlberg:92a} for details),
\begin{IEEEeqnarray*}{rCl}
\hat{\eta}_N^n \to \bar{\eta}^n 
&\defeq & [\bar{R}^n]^{-1} \bar{r}^n,
\end{IEEEeqnarray*}
where $\bar{R}^n$ and $\bar{r}^n$ are the limits of $R^n_N$ and $r^n_N$ w.p.1, respectively.

To guarantee that the true system~\eqref{eq:true-ARX-model} is asymptotically in the model set defined by the ARX model~\eqref{eq:ARXmodel}, $n$ should be allowed to grow to infinity.
Accordingly, we let the model order depend on the sample size $N$. 
For our theoretical results, we use the following assumption.

\begin{assumption}[Model order]
\label{ass:model-order}
It holds that
\begin{IEEEeqnarray*}{rClrCl}
n(N) & \to &  \infty, \quad & N & \to & \infty \\
n(N)^{4+\delta}/N & \to & 0, \quad & N & \to & \infty
\end{IEEEeqnarray*}
for some $\delta > 0$.
\end{assumption}

Introduce the notation $\hat{\eta}_N := \hat{\eta}_N^{n(N)}$
and, for future reference,
\begin{IEEEeqnarray}{rClrCl}
\eta_\nul^n &:=& 
\begin{bmatrix}
{a}_1^\nul & \dotsc & {a}_n^\nul & {b}_1^\nul & \dotsc & {b}_n^\nul
\end{bmatrix} \transpose ,
\\
\eta_\nul &:=& 
\begin{bmatrix}
a_1^\nul & {a}_2^\nul & \dotsc & {b}_1^\nul & {b}_2^\nul & \dotsc 
\end{bmatrix} \transpose .
\end{IEEEeqnarray} 
The asymptotic properties of $\hat{\eta}_N$ have been established in \cite{Ljung&Wahlberg:92a}. We will need the following result on the rate of convergence of the ARX model.

\begin{lemma}
\label{lem:eta_cov}
Assume that Assumptions \ref{ass:true_system}, \ref{ass:input}, \ref{ass:noise} and \ref{ass:model-order} hold. Then with probability 1,
\begin{IEEEeqnarray*}{rClrClrCl}
\sup_\omega \norm{ \begin{bmatrix}
A(\eiw, \hat{\eta}_N)-A^\nul(\eiw) \\
B(\eiw, \hat{\eta}_N)-B^\nul(\eiw)
\end{bmatrix}}_2
&=& \Oc(m(N)),
\end{IEEEeqnarray*}
where
\begin{IEEEeqnarray*}{rClrClrCl}
m(N) &=& n(N) \sqrt{\log N/N} (1+d(N)) + d(N)
\end{IEEEeqnarray*}
and 
\begin{IEEEeqnarray*}{rClrClrCl}
\label{eq:d-def}
 d(N) &\defeq & \sum_{k= n(N)+1}^\infty |a^\nul_k| +  |b^\nul_k| \le \bar C \rho^{n(N)},
 \yesnumber \IEEEeqnarraynumspace
\end{IEEEeqnarray*}
for some $\bar C < \infty$ and $\rho < 1$.
\end{lemma}

\begin{pf}
See Appendix~\ref{app:proof_lem:eta_cov}.
\end{pf}

Lemma~\ref{lem:eta_cov} implies that, as $N$ tends to infinity, the coefficients of $A(q,\hat{\eta}_N)$ converge to those of $A^\nul(q)=1/H^\nul(q)$, and the coefficients of $B(q,\hat{\eta}_N)$ converge to those of $B^\nul(q)=G^\nul(q)/H^\nul(q)$.
Therefore, $B(q,\hat{\eta}_N)/A(q,\hat{\eta}_N)$ can be used as a high order estimate of $G^\nul(q)$, and $1/A(q,\hat{\eta}_N)$ as a high order estimate of $H^\nul(q)$.
We thus define these high order estimates by
\begin{equation}
G(q,\hat{\eta}_N) := \frac{B(q,\hat{\eta}_N)}{A(q,\hat{\eta}_N)} , \quad H(q,\hat{\eta}_N) := \frac{1}{A(q,\hat{\eta}_N)} .
\end{equation}

Despite the simplicity of ARX models, they are not appropriate to model~\eqref{ass:true_system} for most practical uses.
As the order $n$ is required to be arbitrarily large, the estimated model will have unacceptably high variance.

Nevertheless, the high order ARX model estimate can be used to obtain a model of low order, reducing the variance.
This can be done efficiently without re-using the data.
The reason is that the estimate $\hat{\eta}_N$ and its covariance are asymptotically a sufficient statistic for our problem.
To observe this, consider the infinite order ARX model
\begin{equation}
y_t = \varphi_t\transpose \eta + e_t ,
\end{equation}
where
\begin{IEEEeqnarray}{rClrCl}
\varphi_t &:=&
\begin{bmatrix}
-y_{t-1} & -y_{t-2} & \dotsc & u_{t-1} & u_{t-2} & \dotsc 
\end{bmatrix} \transpose , \\
\eta &:=& 
\begin{bmatrix}
a_1 & {a}_2 & \dotsc & {b}_1 & {b}_2 & \dotsc 
\end{bmatrix} \transpose .
\end{IEEEeqnarray} 
Then, the probability density function of $y^N:=\{y_t\}_{t=1}^N$ given $\eta$ is
\begin{equation}
\begin{aligned}
&f(\theta;y^N) = \prod_{t=1}^N \frac{1}{\sqrt{2\pi\sigma_\nul^2}}e^{-\frac{y_t-\varphi_t\transpose \eta}{2\sigma^2_\nul}} \\
&= C e^{-\frac{1}{2\sigma_\nul^2} \left( \eta\transpose \sum_{t=1}^N \varphi_t \varphi_t\transpose \eta + 2 \sum_{t=1}^N \varphi_t\transpose y_t \eta \right)} e^{-\frac{1}{2\sigma_\nul^2} \sum_{t=1}^N y_t^2} .
\end{aligned}
\end{equation}
where we treat $\sigma_\nul^2$ as a known constant (in this case, because it is a scalar, it does not influence the estimation of the parameters of interest).
Then, it follows from~\cite{lehman} that $R_N:=\sum_{t=1}^N \varphi_t \varphi_t\transpose$ and $r_N:=\sum_{t=1}^N \varphi_t\transpose y_t$ form a sufficient statistic for the data $y^N$.
Alternatively, since $\hat{\eta}_N=R_N^{-1} r_N$, we can say that $\hat{\eta}_N$ and $R_N$ are the sufficient statistic. 
However, when $n$ is finite, there is a bias error induced by the truncation of the parameter sequences $\{a_k\}$ and $\{b_k\}$.
If that error is assumed to be small, the estimate $\hat{\eta}^n_N$ will contain practically the same information about the system dynamics as the data. 
If the order $n$ is allowed to tend to infinity as a function of the sample size $N$, according to Assumption~\ref{ass:model-order}, then the estimate $\hat{\eta}_N$ is, asymptotically, a sufficient statistic.
Thus, the data could in principle be disregarded, and $\hat{\eta}_N$ alone be used to obtain an estimate of a lower order model that is asymptotically efficient.

%% file: using-ARX.tex
Having estimated a high order ARX model, we are interested in using this estimate to obtain a low order estimate $G(q,\theta)$.
In this section, we discuss available approaches to do so.

\subsection{Exact Maximum Likelihood}

Being a sufficient statistic, $\hat{\eta}_N$ and its covariance can be used to obtain an estimate of $\theta$ that is asymptotically efficient.
This can be done using an exact ML criterion~\cite{wahlberg89}.
Let $\eta^n(\theta,\alpha)$ be the parameter vector $\eta^n$ obtained from $\theta$ and $\alpha$, satisfying the relations
\begin{equation}
A(q,\eta) = \frac{1}{H(q,\alpha)} , \quad B(q,\eta) = \frac{G(q,\theta)}{H(q,\alpha)} .
\end{equation}
This procedure consists in minimizing 
\begin{equation}
\left[\hat{\eta}_N - \eta^n(\theta,\alpha) \right]^\top \left[ \text{cov}\left(\hat{\eta}_N\right) \right]^{-1} \left[ \hat{\eta}_N - \eta^n(\theta,\alpha) \right] ,
\label{eq:exactML}
\end{equation}
where $\text{cov}\left(\hat{\eta}_N\right)$ denotes the covariance of the estimated vector $\hat{\eta}_N$.
Since this covariance matrix is in general unknown, in practice the cost function~\eqref{eq:exactML} requires an approximation.
We consider two possibilities that do not affect the asymptotic properties of the obtained estimates.

One possibility consists in replacing $\left[\text{cov}\left(\hat{\eta}_N\right)\right]^{-1}$ by a consistent estimate---for example, $R^n_N$~\cite{wahlberg89}.
In this case, we minimize
\begin{equation}
\left[\hat{\eta}_N - \eta^n(\theta,\alpha) \right]^\top R^n_N \left[ \hat{\eta}_N - \eta^n(\theta,\alpha) \right] ,
\label{eq:exactML-consR}
\end{equation}
which yields asymptotically efficient estimates of $G(q,\theta)$ and $H(q,\alpha)$.
Because $\eta^n(\theta,\alpha)$ is nonlinear in general, minimizing~\eqref{eq:exactML-consR} is a non-convex optimization problem.

Another possibility is to write the covariance matrix as function of the low order parameters $\theta$ and $\alpha$---denoted $R^n(\theta,\alpha)$ (see~\cite{wahlberg89} for details).
In this case, we minimize the criterion
\begin{equation}
\left[\hat{\eta}_N - \eta^n(\theta,\alpha) \right]^\top R^n(\theta,\alpha) \left[ \hat{\eta}_N - \eta^n(\theta,\alpha) \right] ,
\label{eq:exactML-Rtheta}
\end{equation}
Although minimizing~\eqref{eq:exactML-Rtheta} seems, at first sight, more complicated than minimizing~\eqref{eq:exactML-consR}, it is observed in~\cite{wahlberg89} that the cost function~\eqref{eq:exactML-Rtheta} can be approximated by an asymptotic ML criterion that allows separating the estimation of $G(q,\theta)$ and $H(q,\alpha)$, while still providing asymptotically efficient estimates.

\subsection{Asymptotic Maximum Likelihood (ASYM)}

As shown in~\cite{wahlberg89}, minimizing~\eqref{eq:exactML-Rtheta} is asymptotically the same as minimizing
\begin{multline}
\int_0^{2\pi} \abs{G(e^{i\omega},\hat{\eta}_N) - G(e^{i\omega},\theta)}^2 \frac{\Phi_u(e^{i\omega})}{\abs{H(e^{i\omega},\hat{\eta}_N)}^2} d\omega \\ +
\frac{\hat{\sigma}^2}{2\pi} \int_0^{2\pi} \frac{\abs{H(e^{i\omega},\hat{\eta}_N) - H(e^{i\omega},\alpha)}^2}{\abs{H(e^{i\omega},\hat{\eta}_N)}^2} d\omega ,
\label{eq:asymML-Rtheta}
\end{multline}
where $\hat{\sigma}^2$ is a consistent estimate of $\sigma^2_\nul$.
Because the first term in~\eqref{eq:asymML-Rtheta} is only dependent on $G(q,\theta)$ and the second term on $H(q,\alpha)$, $G(q,\theta)$ can be estimated by minimizing the first term.
Then, the minimization problem we are interested in becomes
\begin{equation}
V_N(\theta) \! = \! \int_0^{2\pi} \!\! \abs{G(e^{i\omega},\hat{\eta}^n_N) - G(e^{i\omega},\theta)}^2 \frac{\Phi_u(e^{i\omega})}{\abs{H(e^{i\omega},\hat{\eta}_N)}^2} d\omega .
\label{eq:asymML-Rtheta-onlyG}
\end{equation}

The idea of the ASYM method~\cite{zhu_book} is to minimize the time domain equivalent to~\eqref{eq:asymML-Rtheta-onlyG} for finite sample size:
\begin{equation}
V_N(\theta) = \frac{1}{N} \sum_{t=1}^N \left[ \left( \frac{B(q,\hat{\eta}_N)}{A(q,\hat{\eta}_N)} - G(q,\theta) \right) A(q,\hat{\eta}_N) u_t \right]^2 .
\label{eq:VN_asym}
\end{equation}
Minimizing~\eqref{eq:VN_asym} is still a non-convex optimization problem. 
However, it is pointed out in~\cite{zhu_book} that this minimization problem has an advantage over directly estimating $G(q,\theta)$ using PEM, which makes the method numerically more reliable.
Because the output is not used explicitly in~\eqref{eq:VN_asym}, and the noise contribution is only present indirectly through the high order estimates, the influence of the disturbance is reduced.

\subsection{BJSM method}

In alternative to using local non-linear optimization techniques, the BJSM method uses the Steiglitz-McBride iterations.
The idea of BJSM is to first estimate a high order ARX model and then apply the Steiglitz-McBride method~\cite{stmcb_original} to a data set pre-filtered by the ARX model estimate.
The estimates obtained are asymptotically efficient in open loop.
Because BJSM uses the Steiglitz-McBride, we start by reviewing the latter.

\subsubsection{Steiglitz-McBride}

The setting for the Steiglitz-McBride algorithm is when the transfer function $H^\nul(q)$ equals one (\ie $C^\nul(q)=D^\nul(q)=1$).
The objective is to estimate $L(q,\theta)$ and $F(q,\theta)$.

Consider the following three steps.
First, an ARX model
\begin{equation*}
  F(q,\theta) y_t = L(q,\theta) u_t + e_t
\end{equation*}
is estimated using least squares, providing an initilialization estimate $\hat{\theta}_N^0$. 
Second, the output and input are filtered by
\begin{equation*}
    y^f_t = \frac{1}{F(q,\hat{\theta}_N^{1})} y_t , \qquad
    u^f_t = \frac{1}{F(q,\hat{\theta}_N^{1})} u_t .
\end{equation*}
Third, least squares is applied to the ARX model
\begin{equation*}
  F(q,\theta) y^f_t = L(q,\theta) u^f_t + e_t ,
\end{equation*}
providing a new estimate---$\hat{\theta}_N^1$. 
Then, we can continue to iterate by repeating Steps 2 and 3.
We define the estimate obtained at iteration $k$ by $\hat{\theta}^k_N$.

Notice that, since the true system has an OE structure, and we are estimating an ARX model, we are actually minimizing, in Step 1, the function
\begin{IEEEeqnarray}{rCl}
  V_N(\theta) &=& \frac{1}{N} \sum_{t=1}^N \left[F(q,\theta)y_t-L(q,\theta)u_t\right]^2 ,
\label{eq:StMcB_costfunc_step2}
\end{IEEEeqnarray}
which, evaluated at the true parameter $\theta_\nul$, equals
\begin{IEEEeqnarray}{rCl}
  V_N(\theta_\nul) &=& \frac{1}{N} \sum_{t=1}^N \left[ F(q,\theta_\nul) e_t \right]^2 .
\label{eq:StMcB_costfunc_step2_true}
\end{IEEEeqnarray}
From~\eqref{eq:StMcB_costfunc_step2_true}, we observe the true parameter $\theta_\nul$ does not correspond to the cost function of a white sequence. 
Consequently, the initialization estimate $\hat{\theta}^0_N$ is not consistent.
However, at iteration $k$ we have, evaluated at $\theta=\theta_\nul$,
\begin{IEEEeqnarray}{rCl}
  V_N(\theta_\nul) &=& \frac{1}{N} \sum_{t=1}^N \left[ \frac{F(q,\theta_\nul)}{F(q,\hat{\theta}_N^k)} e_t \right]^2 .
\label{eq:StMcB_costfunc_step3}
\end{IEEEeqnarray}
So, assuming convergence to the true parameters (i.e., $\hat{\theta}_N^{k}\to\theta_\nul$, as $k\to\infty$ and $N\to\infty$), \eqref{eq:StMcB_costfunc_step3} asymptotically corresponds to~\eqref{eq:lossfunc} for an OE model structure.

Convergence of the Steiglitz-McBride has been studied in \cite{sm_ss81}, where it is shown that the method is locally convergent when the additive output noise is white.
Moreover, it will be globally convergent if the signal-to-noise ratio is sufficiently large. 
Assuming convergence, the estimates are asymptotically Gaussian distributed. However, in general, the covariance of the estimated parameters does not asymptotically attain $M^{-1}_\text{CR}$.

The Steiglitz-McBride is thus an attempt to minimize~\eqref{eq:lossfunc}, but it only does so consistently with additive white noise, and even then it is not asymptotically efficient.

\subsubsection{BJSM}

In \cite{zhu_hakan}, the Box-Jenkins Steiglitz-McBride (BJSM) algorithm is introduced. 
This algorithm copes with two limitations of the Steiglitz-McBride. 
First, it is consistent for systems with BJ structure, instead of only OE. 
Second, it is asymptotically efficient for open loop data. 

The method uses the following procedure. 
First, an ARX model~\eqref{eq:ARXmodel} is estimated with least squares. 
Second, the original data set is pre-filtered by $A(q,\hat{\eta}_N)$.
Third, the Steiglitz-McBride algorithm is applied to the pre-filtered data set. 

Recall that, to be convergent, the Steiglitz-McBride algorithm requires that $H^\nul(q)=1$. 
The main idea of BJSM is thus to use $A(q,\hat{\eta}_N)$ as an estimate of $[H^\nul(q)]^{-1}$ and pre-filter the data according to
\begin{IEEEeqnarray*}{rClrCl}
y_t^{\text{pf}} &=& A(q,\hat{\eta}_N) y_t, & \qquad u_t^{\text{pf}} &=& A(q,\hat{\eta}_N) u_t .
\label{eq:oldfilter}
\end{IEEEeqnarray*}
Then, the pre-filtered data satisfies
\begin{IEEEeqnarray}{rClrCl}
y_t^{\text{pf}} &=& \frac{L^\nul(q)}{F^\nul(q)} u_t^{\text{pf}} + A(q,\hat{\eta}_N) H^\nul(q) e_t ,
\label{eq:oldbjsm_eq}
\end{IEEEeqnarray}
which asymptotically is according to, due to Lemma~\ref{lem:eta_cov},
\begin{IEEEeqnarray}{rClrCl}
y_t^{\text{pf}} &\approx& \frac{L^\nul(q)}{F^\nul(q)} u_t^{\text{pf}} + e_t .
\label{eq:oldbjsm_eq_approx}
\end{IEEEeqnarray}
Since~\eqref{eq:oldbjsm_eq_approx} is of OE structure, the Steiglitz-McBride algorithm can be applied to the data set $\{y_t^{\text{pf}},u_t^{\text{pf}}\}$.

Notice that, if we were to apply PEM to the pre-filtered data set, we would minimize, motivated by~\eqref{eq:oldbjsm_eq_approx},
\begin{IEEEeqnarray}{rCl}
  V_N(\theta) &=& \frac{1}{N} \sum_{t=1}^N \left( y_t^{\text{pf}} - \frac{L(q,\theta)}{F(q,\theta)} u_t^{\text{pf}} \right)^2 .
\label{eq:BJSM_cost_func}
\end{IEEEeqnarray}
To avoid an explicit non-convex minimization problem, we use the Steiglitz-McBride method instead.
Although the Steiglitz-McBride is not asymptotically efficient, the BJSM method is when used with open loop data~\cite{zhu_hakan}.

However, not all the information in $\hat{\eta}_N$ is being used, as the filtering~\eqref{eq:oldfilter} only uses $A(q,\hat{\eta}_N)$.
In other words, the ARX model is not used as a sufficient statistic for this problem.
For the method to still be asymptotically efficient, the output data are used when constructing the pre-filtering.
This leads to two limitations.

The first is a counter-intuitive result. 
Suppose that $H^\nul(q)=1$ (i.e., the true system is already of OE structure).
Then, we have that $A^\nul(q)=1$, and estimating a finite impulse response (FIR) model would suffice to asymptotically model the true system.
However, this would maintain the data set unchanged when applying the filtering~\eqref{eq:oldfilter}, and BJSM would simply be reduced to the Steiglitz-McBride method, which is not asymptotically efficient.
If, on the other hand, it is not assumed that ${A^\nul(q)=1}$ and an estimate $A(q,\hat{\eta}_N)$ is still computed, BJSM will be asymptotically efficient.
Thus, although an FIR model is asymptotically a sufficient statistic for a system of OE structure (like the ARX model is for BJ structures) it is not possible to make use of this information when applying the BJSM method, since it does not exploit the full statistical properties of the high order model.

As for the second limitation, we observe that although BJSM avoids solving a non-convex optimization problem by applying the Steiglitz-McBride algorithm, it has the disadvantage of requiring the number of iterations of the Steiglitz-McBride to tend to infinity in order to provide consistent and asymptotically efficient estimates~\cite{zhu_hakan}.
To bypass this problem but still avoid a non-convex minimization procedure, we use the Steiglitz-McBride with the ASYM method.
This will allow us to obtain an asymptotically efficient estimate in one iteration.

% \subsection{Summary}

% In summary, the ASYM method proposes a procedure to obtain an estimate of $G(q,\theta)$ from the high order ARX model estimate.
% This is based on an asymptotic ML criterion.
% For this method, a non-convex optimization procedure is required, and no theoretical results regarding asymptotic covariance are available.

% Concerning the BJSM method, it provides an estimate of $G(q,\theta)$ that is asymptotically efficient in open loop, avoiding the non-convexity of PEM and a choice for the model order of the noise model.
% However, it is still an iterative method, as it relies on convergence of the Steiglitz-McBride iterations.
% Moreover, it still requires a high order ARX model even if the noise contribution is white, in which case a high order FIR should be sufficient.

% To understand the limitations of these methods, we will take a step back in the next section, and consider an exact maximum likelihood criterion to obtain $\theta$ from the high order ARX model estimate.

%% file: method.tex
The objective of our approach is to minimize~\eqref{eq:VN_asym} without using a non-convex optimization method.
To do so, we use an approach that combines ideas from ASYM and BJSM.

First, we write~\eqref{eq:VN_asym} as
\begin{equation}
V_N(\theta) = \frac{1}{N} \sum_{t=1}^N \left[ B(q,\hat{\eta}_N) u_t - \frac{L(q,\theta)}{F(q,\theta)} A(q,\hat{\eta}_N) u_t \right]^2 .
\label{eq:newVN}
\end{equation}
Then, we notice that~\eqref{eq:newVN} has the same form as~\eqref{eq:BJSM_cost_func} if we define
\begin{equation}
y^{\text{pf}}_t := B(q,\hat{\eta}_N) u_t , \qquad u^{\text{pf}}_t := A(q,\hat{\eta}_N) u_t ,
\label{eq:newpf}
\end{equation}
and thus the same idea (i.e., applying the Steiglitz-McBride to $\{y^{\text{pf}}_t,u^{\text{pf}}_t\}$) can be used.

The only difference between this approach and BJSM is in the pre-filtered output.
Comparing~\eqref{eq:newpf} and~\eqref{eq:oldfilter}, we observe that $u^{\text{pf}}_t$ are defined similarly, but $y^{\text{pf}}_t$ are different.
The difference lies in the true output not being used to construct the new pre-filtered data set.
Rather, it is simulated from the input and the ARX model estimate.
Indeed, we can simulate the output with
\begin{IEEEeqnarray}{rCl}
  y_t^s & \defeq & \frac{B(q,\hat{\eta}_N)}{A(q,\hat{\eta}_N)} u_t ,
\end{IEEEeqnarray}
and then apply the same filter as in~\eqref{eq:oldfilter}, but on the simulated output, according to
\begin{IEEEeqnarray}{rCl}
y_t^{\text{pf}} &=& A(q,\hat{\eta}_N) y_t^s = B(q,\hat{\eta}_N) u_t ,
\IEEEeqnarraynumspace
\end{IEEEeqnarray}
obtaining the proposed pre-filter~\eqref{eq:newpf}.

In summary, the proposed method is as follows:
\begin{enumerate}[(1)]
\item estimate an ARX model using the input-output data $\{u_t,y_t\}$, $t = 1, \dotsc, N$, according to \eqref{eq:eta_N-def};
\item construct the pre-filtered data $\{u^{\text{pf}}_t,y^{\text{pf}}_t\}$, according to~\eqref{eq:newpf};
\item apply the Steiglitz-McBride method with $\{u^{\text{pf}}_t,y^{\text{pf}}_t\}$ to obtain estimates $L(q,\hat{\theta}_N)$ and $F(q,\hat{\theta}_N)$ of $L^\nul(q)$ and $F^\nul(q)$, respectively.
\end{enumerate}

Note that the pre-filtered data set~\eqref{eq:newpf} only depends on the original output data $\{y_t\}$ through the least squares estimate $\hat{\eta}_N$. 
With this method, we use the high order ARX model as an asymptotic sufficient statistic for our problem, and disregard the data without loss of information.
Indeed, as will be shown in the next section, this procedure is asymptotically efficient for open loop data.

Moreover, there are two advantages for disregarding the data after the high order ARX model has been estimated.
Although these are formally shown in the next section, we observe them here, supported by intuitive explanations.

First, the pre-filter~\eqref{eq:newpf} uses the complete statistical information contained in the estimate $\hat{\eta}_N$. 
So, if the noise contribution affecting the true system~\eqref{eq:truesys} is white, a high-order FIR model can be estimated instead of an ARX.
In this case, $A(q,\hat{\eta}_N)=1$.

Second, this procedure asymptotically (in $N$) only requires one iteration.
To intuitively understand why this is the case, we recall why the Steiglitz-McBride is an iterative method.
Note that the initialization step of the Steiglitz-McBride minimizes~\eqref{eq:StMcB_costfunc_step2}, which, when evaluated at the true parameters, as in~\eqref{eq:StMcB_costfunc_step2_true}, does not correspond to a cost function of white sequence.
Therefore, the initialization estimate $\hat{\theta}^0_N$ is biased.
Then, we start iterating.
At the first iteration, the cost function evaluated at $\theta_\nul$ is given by~\eqref{eq:StMcB_costfunc_step3} with $F(q,\hat{\theta}^{0}_N)$. 
Because $F(q,\hat{\theta}^{0}_N)$ is biased, the true parameter will still not correspond to the cost function of a white sequence.
Therefore, the new estimate $\hat{\theta}^1_N$ will not be consistent either.
However, by continuing to iterate, it can be shown that, under the conditions observed in~\cite{stoica:00}, $\hat{\theta}^{k}_N\rightarrow\theta_\nul$, as $k\rightarrow\infty$ and $N\to\infty$.
Concerning the original BJSM method, since the pre-filtered data is according to~\eqref{eq:oldbjsm_eq}, it is asymptotically approximately an OE model structure, and a similar procedure takes place.

On the other hand, the alternative pre-filtering, which disregards the original data, satisfies
\begin{IEEEeqnarray}{rClrCl}
y_t^{\text{pf}} &=& \frac{L^\nul(q)}{F^\nul(q)} u_t^{\text{pf}} + \left( \frac{B(q,\hat{\eta}_N)}{A(q,\hat{\eta}_N)} - \frac{L^\nul(q)}{F^\nul(q)} \right) u_t^{\text{pf}} .
\label{eq:newfiltereq}
\end{IEEEeqnarray}
This is a noise-free equation, except for the noisy parameters in the ARX model. 
However, from Lemma~\ref{lem:eta_cov}, the second term in~\eqref{eq:newfiltereq} tends to zero asymptotically.
As consequence, the variance of the error sequence being minimized by the Steiglitz-McBride iterations disappears asymptotically, and only one iteration is required.

We observe that the proposed method essentially consists of applying the Steiglitz-McBride algorithm to perform model order reduction based on an asymptotic ML criterion.
We will thus refer to the method as Model Order Reduction Steiglitz-McBride (MORSM).
The idea of using the Steiglitz-McBride to, in some sense, perform model order reduction, is not new.
Variants of the Steiglitz-McBride method have been applied to estimate rational filters from an impulse response estimate, instead of applying the method directly to data (see, \eg \cite{evansfischl73,mcclellanlee91,shaw94}).
However, although some of these procedures are in some sense optimal under specific conditions, we consider a quite general system identification problem and motivate the application of the method based on an ML criterion.
This, as we proceed to show, not only provides asymptotically efficient estimates under a quite general class of systems and external signals, but also does so in one iteration.

%% file: properties.tex
In this section, we analyze both the convergence and asymptotic covariance of the proposed method.
To derive these results, we will need a formal expression for the estimate of $\theta$ at iteration $k+1$ of the MORSM algorithm.
Define
\begin{IEEEeqnarray*}{rClrCl}
y_t(\eta,\theta) &=& \frac{B(q, \eta)}{F(q,\theta)}u_t, \quad & y_t(\eta_\nul,\theta) &=& \frac{B^\nul(q)}{F(q,\theta)}u_t, \IEEEeqnarraynumspace
\\
u_t(\eta,\theta) &=& \frac{A(q,\eta)}{F(q,\theta)}u_t, \quad & y_t(\eta_\nul,\theta) &=& \frac{A^\nul(q)}{F(q,\theta)}u_t,
\end{IEEEeqnarray*}
and
\begin{IEEEeqnarray*}{rClrCl}
\xi_t(\eta,\theta) &=& \frac{L^\nul(q)}{F^\nul(q)} \frac{B(q,\eta) - B^\nul(q)}{B^\nul(q)} u_t(\eta,\theta)
\\ &&
-  \frac{A(q,\eta) - A^\nul(q) }{A^\nul(q)} y_t(\eta,\theta).
\end{IEEEeqnarray*}
The same definition also applies to vector valued signals, such as \eqref{eq:varphi-def}.

Using that the pre-filtered data set consists of filtered versions of $u_t$ and that $G(q)$ can be represented both using $L^\nul(q)$ and $F^\nul(q)$ as well as using $B^\nul(q)$ and $A^\nul(q)$,
we have that
\begin{IEEEeqnarray}{rCcCl}
\label{eq:exact-model}
u_t &=& \frac{1}{B(q,\hat \eta_N)}  y_t^{\text{pf}} &=& 
\frac{L^\nul(q)A^\nul(q)}{F^\nul(q)B^\nul(q)}\frac{1}{A(q,\hat \eta_N)} u_t^{\text{pf}}.
\end{IEEEeqnarray}
Filtering \eqref{eq:exact-model} by
\begin{IEEEeqnarray*}{rClrCl}
\label{eq:exact-model2}
F^\nul(q) \frac{A(q,\hat \eta_N)B(q,\hat \eta_N)}{A^\nul(q)F(q, \hat \eta_N) },
\end{IEEEeqnarray*}
we arrive at the noise-free equation
\begin{IEEEeqnarray*}{rClrCl}
\label{eq:exact-model2}
F^\nul(q)\frac{A(q,\hat{\eta}_N)}{A^\nul(q)}  y_t(\hat \eta_N,\hat{\theta}^k_N) &=& L^\nul(q)\frac{B(q,\hat{\eta}_N) }{B^\nul(q)} u_t(\hat \eta_N,\hat{\theta}^k_N)
\end{IEEEeqnarray*}
relating the pre-filtered data.
Equivalently,
\begin{IEEEeqnarray*}{rClrCl}
 F^\nul(q) y_t(\hat{\eta}_N,\hat{\theta}^k_N) &=& L^\nul(q)  u_t(\hat{\eta}_N,\hat{\theta}^k_N)  + F^\nul(q)  \xi_t(\hat{\eta}_N,\hat{\theta}^k_N) ,
\end{IEEEeqnarray*}
which can be written in regression form as
\begin{IEEEeqnarray}{rClrCl}
\label{eq:regression-form}
 y_t(\hat{\eta}_N,\hat{\theta}^k_N) &=&
 [\varphi^m(\hat{\eta}_N , \hat{\theta}^k_N)]\transpose \theta_\nul + F^\nul(q)  \xi_t(\hat{\eta}_N,\hat{\theta}^k_N) .   \IEEEeqnarraynumspace
 \label{eq:linear-regression}
\end{IEEEeqnarray}
Given $\hat{\theta}_N^k$, the next parameter estimate in the Steiglitz-McBride iterations $\hat{\theta}_N^{k+1}$, is defined as the least squares estimate of $\theta_\nul$ in the linear regression \eqref{eq:regression-form}:
\begin{IEEEeqnarray}{rClrCl}
\label{eq:theta_update-equation}
 \hat{\theta}^{k+1}_N &=&
 [R^m(\hat{\eta}_N , \hat{\theta}^k_N)]^{-1} r^m(\hat{\eta}_N,\hat{\theta}^k_N),
\end{IEEEeqnarray}
where
\begin{IEEEeqnarray*}{rClrCl}
R^m(\eta^n , \theta) &=& \frac{1}{N} \sum_{t = m+1}^N \varphi_t^m(\eta^n,\theta)(\varphi_t^m(\eta^n,\theta))\transpose, \\
r^m(\eta^n , \theta) &=& \frac{1}{N} \sum_{t = m+1}^N \varphi_t^m(\eta^n,\theta)y_t(\eta^n,\theta).
\end{IEEEeqnarray*}

Notice that \eqref{eq:linear-regression} is a linear regression form of \eqref{eq:newfiltereq} with the notable difference that the error made in the ARX model enters linearly into $\xi_t(\hat{\eta}_N,\hat{\theta}^k_N)$. As before, the ARX model error tends to zero asymptotically. This is, in essence, what enables the following results.

\begin{theorem}
\label{thm:convergence}
Let Assumptions~\ref{ass:true_system}, \ref{ass:input}, \ref{ass:noise}, and \ref{ass:model-order} hold.
Then, 
\begin{IEEEeqnarray*}{rCl}
 \hat{\theta}^{k}_N & \to & \theta_\nul
 \quad \mathrm{ as \: } N \to \infty , \mathrm{ \: w.p. \: 1, } \quad \mathrm{for \: all \:} k\ge 0
\end{IEEEeqnarray*}
\end{theorem}

\begin{pf}
See Appendix~\ref{sec:proof-of-convergence}.
\end{pf}

Theorem~\ref{thm:convergence} implies that the proposed algorithm achieves consistency in the initialization estimate---that is, $\hat \theta_N^0$ is a consistent estimate of $\theta_\nul$. 
This was not the case for the BJSM algorithm. 

For the asymptotic covariance, we have the following theorem.

\begin{theorem}
\label{thm:asymptotic-covariance}
Let Assumptions~\ref{ass:true_system}, \ref{ass:input}, \ref{ass:noise}, and \ref{ass:model-order} hold. 
Then,
\begin{IEEEeqnarray*}{rClrCl}
\lim_{N \to \infty} N \expect{ ( \hat{\theta}_N^k - \theta_ \nul) ( \hat{\theta}_N^k - \theta_ \nul)\transpose } & = &
\sigma_\nul^2 M_{CR}^{-1},
\end{IEEEeqnarray*}
and $\sqrt{N}( \hat{\theta}_N^k - \theta_ \nul)  \sim \AsN(0,\sigma_\nul^2 M_{CR}^{-1})$ for $k \ge 1$, where $\mathrm{N}$ stands for the normal distribution.
\end{theorem}

\begin{pf}
See Appendix~\ref{sec:a-not-so-long-proof}.
\end{pf}

From Theorem~\ref{thm:asymptotic-covariance}, we observe that the proposed method has the same asymptotic covariance as PEM with Gaussian noise~\eqref{eq:PEMcov}.
Therefore, it is asymptotically efficient with open loop data.
Moreover, the asymptotic efficient estimate is obtained in one iteration, at $k=1$.

%% file: simulation.tex
In this section, we perform two Monte Carlo simulations to study the performance of the method. 
First, we illustrate how it converges in one iteration of the Steiglitz-McBride, while BJSM does not.
Then, we perform a study with random systems, and observe that the method often has better finite sample convergence properties than PEM.

\subsection{One iteration scheme}
\label{subsec:1itsim}
In the first simulation, we compare MORSM and BJSM. 
The practical difference between these methods is in the pre-filtering only. 
In particular, MORSM does not use the noisy output to construct the pre-filtered data set.
The consequence is that the method provides asymptotically efficient estimates in one iteration.

For the simulation, the data are generated by
\begin{equation}
y_t = \frac{q^{-1}+0.1 q^{-2}}{1-1.2q^{-1}+0.6q^{-2}} u_t + \frac{1+0.7q^{-1}}{1-0.9q^{-1}} e_t .
\end{equation}
One hundred Monte Carlo simulations are performed with eight sample sizes logarithmically spaced between $N=200$ and $N=20000$. 
The sequence $\{u_t\}$ is obtained by
\begin{equation}
u_t = \frac{1}{1-q^{-1}+0.89q^{-2}} w_t ,
\end{equation}
where $\{w_t\}$ and $\{e_t\}$ are independent Gaussian white sequences with unit variance.

We compare PEM, BJSM (one and 100 iterations), and MORSM (one and 100 iterations).
All methods estimate a plant parameterized with the correct orders, and PEM also estimates a correctly parameterized noise model.
For BJSM and the proposed method, an ARX model of order 50 is estimated in the first step.
As the objective of this simulation is to observe convergence and asymptotic variance properties, PEM is started at the true parameters, and all methods assume known initial conditions.

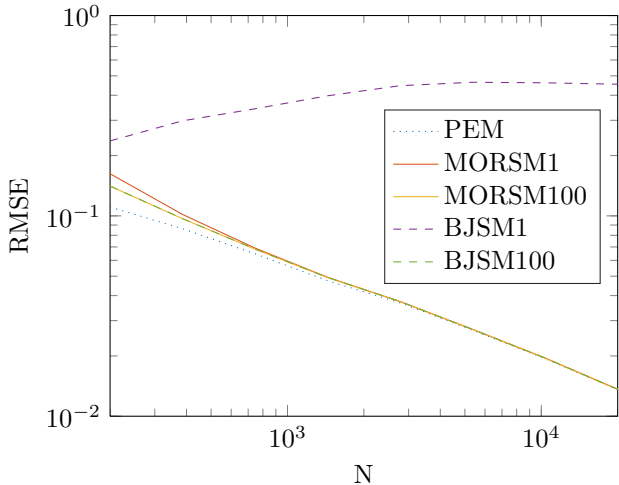
\begin{figure}
\input{sim_convergence_bjsm}
\caption{Average RMSE as function of sample size for several methods, obtained from 100 Monte Carlo runs with a fixed system.}
\label{fig:sim_convergence_bjsm}
\end{figure}

The results are presented in Fig.~\ref{fig:sim_convergence_bjsm}, where the average root mean square error (RMSE) of the impulse response from 1000 Monte Carlo runs is presented for each sample size. 
The RMSE is given by
\begin{equation}
  \text{RMSE} = \norm{g^\nul-\hat{g}}_2,
\end{equation}
where $g^\nul$ is the impulse response of $G^\nul(q)$ and $\hat{g}$ is the impulse response of the estimated plant model.
In Fig.~\ref{fig:sim_convergence_bjsm}, we observe that MORSM and BJSM perform similarly with 100 iterations for all the sample size range used.
MORSM performs slightly worse with one iteration than with 100 for small sample sizes, but they have the same performance for larger $N$.
However, the same is not true for BJSM with one iteration, for which the RMSE does not even decrease with increasing sample size.

In conclusion, if a sufficiently amount of iterations are performed, both MORSM and BJSM attain the asymptotic covariance of PEM.
However, BJSM theoretically needs the Steiglitz-McBride iterations to tend to infinity, while MORSM only needs one iteration.

\subsection{Comparison with PEM}

According to a prediction error criterion, the best model is the one that minimizes a cost function of the prediction errors.
The estimate corresponding to the minimizer of this cost function is asymptotically efficient, meaning that is has asymptotically the minimum possible covariance for a consistent estimator.
The limitation is that this cost function is, in general, non-convex.
Seeking the global minimum requires local non-linear optimization techniques, and it is not guaranteed to be found.
As the number of parameters to estimate increases, PEM has increasingly more difficulty in finding the global minimum.

In the following simulation, we will compare the performance of PEM and the proposed method with randomly generated systems, with structure
\begin{multline}
y_t = \frac{l_1^\nul q^{-1} + l_2^\nul q^{-2} + l_3^\nul q^{-3} + l_4^\nul q^{-4}}{1+f_1^\nul q^{-1} + f_2^\nul q^{-2} + f_3^\nul q^{-3} + f_4^\nul q^{-4}} u_t \\ + \frac{1+c_1^\nul q^{-1}+c_2^\nul q^{-2} + c_3^\nul q^{-3} + c_4^\nul q^{-4}}{1+d_1^\nul q^{-1}+d_2^\nul q^{-2} + d_3^\nul q^{-3} + d_4^\nul q^{-4}} e_t ,
\end{multline}
where $\{u_t\}$ is given as in the previous simulation, and $\{e_t\}$ is Gaussian white noise with variance chosen to obtain a signal-to-noise ratio
\begin{equation}
\text{SNR} =  \frac{\sum_{t=1}^N (u_t)^2}{\sum_{t=1}^N (H^\nul(q)e_t)^2} = 10.
\end{equation}
The coefficients of $L^\nul(q)$ are generated from a uniform distribution, with values between $-1$ and $1$.
The coefficients of the remaining polynomials are generated such that $F^\nul(q)$, $C^\nul(q)$, and $D^\nul(q)$ have all roots inside a half-ring in the unit disc with a radius between $0.7$ and $0.9$, with positive real part.
We do this with the objective of studying a particular class of systems: namely, the systems are effectively of fourth order (i.e., no poles are considerably dominant over others), they can be approximated by ARX models roughly of orders between 30 and 100, and they resemble physical systems.

% \begin{figure}
% \input{Gbodes}
% \caption{Magnitude Bode plots of all randomly generated systems $G^\nul(q)$.}
% \label{fig:bodes}
% \end{figure}

An important practical aspect in implementing the proposed method is how to choose the ARX model order, in case we do not previously have information of an appropriate order to choose.
As we have seen, theoretically the ARX model order should tend to infinity as function of the sample size.
However, for practical purposes it is sufficient to choose an order that can correctly capture the dynamics of the true system.
We then propose the following procedure to choose the order of the ARX model.
Since our objective is to minimize the loss function \eqref{eq:lossfunc} using an indirect approach, we repeat the estimation for a grid of ARX model orders, and choose the low order model that minimizes~\eqref{eq:lossfunc}. 
Since we do not compute a low order noise model, the highest order ARX polynomial $A(q,\hat{\eta}_N)$ is used instead of $1/H(q,\alpha)$ when computing this loss function. 
Although this is a very noisy estimate, the error induced will be the same for every computation, and should not have a considerable influence in choosing the best model.
For the class of systems we consider, we choose the ARX model order from a grid of values between 25 and 125, spaced with intervals of 25.

Moreover, when more than one iteration is used, the same criterion can be applied to optimize over the number of iterations---that is, we choose the model obtained at the iteration that minimizes the cost function~\eqref{eq:lossfunc}.

We compare the following methods:
\begin{itemize}
\item the prediction error method, initialized at the true parameters (PEM true);
\item the prediction error method, initialized with the standard MATLAB procedure (PEM);
\item the Box-Jenkins Steiglitz-McBride method, with 20 iterations (BJSM20);
\item the Model Order Reduction Steiglitz-McBride method, with 20 iterations (MORSM20);
\item the Model Order Reduction Steiglitz-McBride, with one iteration (MORSM1).
\end{itemize}
PEM stops with a maximum of 1000 iterations and a function tolerance of $10^{-5}$, and estimates initial conditions.
MORSM and BJSM truncate initial conditions.
Note that a procedure to estimate initial conditions for this type of methods has been proposed in~\cite{Galrinho:15}, but it is only applicable if the plant and noise model share the same poles (e.g., ARMA, ARMAX) or if the noise model poles are known (e.g., OE), which is not the case of BJ models.

The performance of each method is evaluated by calculating the FIT of the impulse response of the plant, given by, in percent,
\begin{equation}
  \text{FIT} = 100\left(1 - \frac{\text{RMSE}}{\norm{g^\nul-\bar{g}^o}}\right),
\end{equation}
where $\bar{g}^o$ is the average of $g^\nul$.

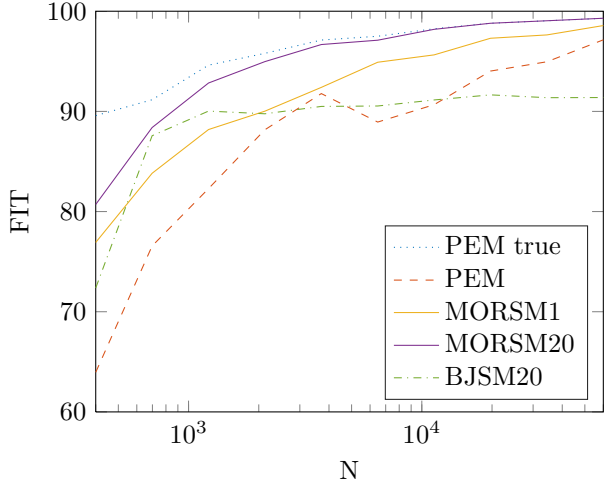
\begin{figure}
\input{sim_random}
\caption{Average FIT for several methods, obtained from 100 Monte Carlo runs with random systems.}
\label{fig:sim_random}
\end{figure}

The results are presented in Fig.~\ref{fig:sim_random}, with the average FIT as function of sample size.
We assume that PEM, when initialized at the true parameters, converges to the global optimum.
Comparing PEM initialized at the true parameters and with the standard MATLAB procedure, we conclude that the latter must sometimes fail to reach the global optimum.

With 20 iterations, MORSM does not seem to reach the global minimum of the prediction error cost function for small sample sizes. 
However, for sample sizes around 4000 and larger, this minimum seems to be attained since MORSM performs similar to PEM initialized at the true parameters. 
This suggests that MORSM may be a viable alternative to PEM when PEM has difficulty in finding the global minimum.

With only one iteration, MORSM performs worse than with 20 iterations for the range of sample sizes used, but their performances becomes closer as larger sample sizes are used.
Theoretically, we have shown that MORSM only requires one iteration to provide asymptotically efficient estimates.
However, we observe that in practice (i.e., for finite sample size), MORSM performs better with more iterations, comparing MORSM1 and MORSM20 in Fig.~\ref{fig:sim_random}.
The fact that in practice MORSM requires more than one iteration to converge does not render it irrelevant in comparison to BJSM.
As we observe in Fig.~\ref{fig:sim_random}, BJSM with 20 iterations does not attain the same asymptotic performance of MORSM because 20 iterations do not seem to be sufficient for BJSM to converge in this simulation, while they are sufficient for MORSM.

%% file: sim_convergence_bjsm.tex
\definecolor{mycolor1}{rgb}{0.00000,0.44700,0.74100}%
\definecolor{mycolor2}{rgb}{0.85000,0.32500,0.09800}%
\definecolor{mycolor3}{rgb}{0.92900,0.69400,0.12500}%
\definecolor{mycolor4}{rgb}{0.49400,0.18400,0.55600}%
\definecolor{mycolor5}{rgb}{0.46600,0.67400,0.18800}%
\begin{tikzpicture}

\begin{axis}[%
width=.8\linewidth,
height=.79*.8\linewidth,
at={(0.758in,0.481in)},
scale only axis,
xmode=log,
xmin=200,
xmax=20000,
xminorticks=true,
xlabel={N},
ymode=log,
ymin=0.01,
ymax=1,
yminorticks=true,
ylabel={RMSE},
axis background/.style={fill=white},
legend style={at={(0.54,0.33)},anchor=south west,legend cell align=left,align=left,draw=white!15!black}
]
\addplot [color=mycolor1,dotted]
  table[row sep=crcr]{%
200	0.110728333126176\\
386	0.0862626630345422\\
746	0.0646363871424342\\
1439	0.0474805202744888\\
2779	0.0368195731463754\\
5365	0.026882354863269\\
10359	0.0193618338275212\\
20000	0.0135782047047229\\
};
\addlegendentry{PEM};

\addplot [color=mycolor2,solid]
  table[row sep=crcr]{%
200	0.162310027397195\\
386	0.101658971540132\\
746	0.0690384178343415\\
1439	0.0493825886212138\\
2779	0.0373516772702931\\
5365	0.0271305047566224\\
10359	0.0195125908072392\\
20000	0.0136330741326931\\
};
\addlegendentry{MORSM1};

\addplot [color=mycolor3,solid]
  table[row sep=crcr]{%
200	0.140786682151778\\
386	0.0968301973005657\\
746	0.0681928809263999\\
1439	0.0492431071101069\\
2779	0.0373400916899822\\
5365	0.0271277094840981\\
10359	0.0195024222083479\\
20000	0.0136285451841922\\
};
\addlegendentry{MORSM100};

\addplot [color=mycolor4,dashed]
  table[row sep=crcr]{%
200	0.236697374156541\\
386	0.298225243980769\\
746	0.342614460478998\\
1439	0.397690315230813\\
2779	0.446587206665001\\
5365	0.464489341180988\\
10359	0.46157571290771\\
20000	0.454209168174153\\
};
\addlegendentry{BJSM1};

\addplot [color=mycolor5,dashed]
  table[row sep=crcr]{%
200	0.140897910850411\\
386	0.0969085965075874\\
746	0.0682624522659668\\
1439	0.0492760303990388\\
2779	0.0373765380796735\\
5365	0.0271365024554734\\
10359	0.0195055811878659\\
20000	0.0136325071057657\\
};
\addlegendentry{BJSM100};

\end{axis}
\end{tikzpicture}%

%% file: sim_random.tex
\definecolor{mycolor1}{rgb}{0.00000,0.44700,0.74100}%
\definecolor{mycolor2}{rgb}{0.85000,0.32500,0.09800}%
\definecolor{mycolor3}{rgb}{0.92900,0.69400,0.12500}%
\definecolor{mycolor4}{rgb}{0.49400,0.18400,0.55600}%
\definecolor{mycolor5}{rgb}{0.46600,0.67400,0.18800}%
\begin{tikzpicture}

\begin{axis}[%
width=.8\linewidth,
height=.79*.8\linewidth,
at={(0.758in,0.481in)},
scale only axis,
xmode=log,
xmin=400,
xmax=60000,
xminorticks=true,
xlabel={N},
ymin=60,
ymax=100,
ylabel={FIT},
axis background/.style={fill=white},
legend style={at={(0.97,0.03)},anchor=south east,legend cell align=left,align=left,draw=white!15!black}
]
\addplot [color=mycolor1,dotted]
  table[row sep=crcr]{%
400	89.5792088891486\\
698	91.1636702978664\\
1218	94.6163586121497\\
2125	95.7953506093155\\
3709	97.1232452240253\\
6471	97.5015936373115\\
11292	98.2497127605789\\
19705	98.7992716768252\\
34385	99.0474989147028\\
60000	99.2911690375527\\
};
\addlegendentry{PEM true};

\addplot [color=mycolor2,dashed]
  table[row sep=crcr]{%
400	63.9537835374271\\
698	76.5652149066319\\
1218	82.2906773993303\\
2125	88.1670176640462\\
3709	91.7784926672282\\
6471	88.9353902272836\\
11292	90.6689003891757\\
19705	94.0262217304833\\
34385	94.9545652501671\\
60000	97.1445548718658\\
};
\addlegendentry{PEM};

\addplot [color=mycolor3,solid]
  table[row sep=crcr]{%
400	76.9200002681013\\
698	83.8222976545117\\
1218	88.1880701179329\\
2125	90.0235951711768\\
3709	92.397588496183\\
6471	94.9072004604888\\
11292	95.6424722998808\\
19705	97.3013387475649\\
34385	97.6348150729943\\
60000	98.5693271543043\\
};
\addlegendentry{MORSM1};

\addplot [color=mycolor4,solid]
  table[row sep=crcr]{%
400	80.7118293818219\\
698	88.387976876188\\
1218	92.8338632914309\\
2125	94.9707309488114\\
3709	96.6746639871877\\
6471	97.1072452785198\\
11292	98.1933185057791\\
19705	98.7985988118054\\
34385	99.0488568767059\\
60000	99.2980108096628\\
};
\addlegendentry{MORSM20};

\addplot [color=mycolor5,dashdotted]
  table[row sep=crcr]{%
400	72.3712891755974\\
698	87.5708624913756\\
1218	90.0312596882531\\
2125	89.7619712040688\\
3709	90.4967113925102\\
6471	90.539649339037\\
11292	91.1417547889768\\
19705	91.6501160196333\\
34385	91.3692389323208\\
60000	91.3836817009678\\
};
\addlegendentry{BJSM20};

\end{axis}
\end{tikzpicture}%

%% file: appendices.tex
\section{Proof of Lemma~\ref{lem:eta_cov}}
\label{app:proof_lem:eta_cov}

%\ref{ass:true_system}, \ref{ass:input}, \ref{ass:noise} and \ref{ass:model-order}
The result follows from Theorem 3.1 in \cite{Ljung&Wahlberg:92a}. 
Next, we verify the conditions of that theorem. 
Assumption~\ref{ass:true_system} and the finite dimensionality of $G^\nul$ and $H^\nul$ implies that
\begin{IEEEeqnarray}{rCl}
\label{eq:bound_a_k_b_k}
\max(|a_k|, |b_k|) \le C\rho^k
\end{IEEEeqnarray}
for some $C < \infty$ and $0 < \rho < 1$. 
This implies that Condition S1 holds. Furthermore, the bound \eqref{eq:bound_a_k_b_k} implies the inequality in \eqref{eq:d-def} for some $\bar C < ̃\infty$. 
Assumption~\ref{ass:noise} clearly implies Condition S2 (for any $p \le 5$). 
Assumption~\ref{ass:model-order} implies Conditions D1 and D3.
Thus all conditions in Theorem 3.1 of \cite{Ljung&Wahlberg:92a} have been verified
and the result in the lemma follows from this theorem.

\section{Proof of Theorem~\ref{thm:convergence}}
\label{sec:proof-of-convergence}

Using Parseval's formula, we have
\begin{IEEEeqnarray}{rClrCl}
\label{eq:R_bar_w}
 \bar{R}(\theta) &=& \frac{1}{2\pi}\wint{
 \:
 \begin{bmatrix}
 -B^\nul \Gamma_m \\
 A^\nul \Gamma_m
 \end{bmatrix}
 \begin{bmatrix}
 -B^\nul \Gamma_m \\
 A^\nul \Gamma_m
 \end{bmatrix}^*
 \frac{\Phi_u}{|F(\theta)|^2}
 } \IEEEeqnarraynumspace
\end{IEEEeqnarray}
We notice that $\bar{R}(\theta) > 0$ whenever $\theta$ is in the stability region for the coefficients of polynomials of degree $m$
\begin{IEEEeqnarray}{rClrCl}
\bar S \defeq \{ \theta : F(z,\theta) = 0 \Rightarrow |z| < 1 \} \subset \Rb^{2m}
\end{IEEEeqnarray}
We introduce the notation
\begin{IEEEeqnarray*}{rCl}
 f(N) &=& \Oc(g(N))
\end{IEEEeqnarray*}
to mean that $f(N)$ decays to zero with the rate $g(N)$, i.e., that there exists some positive constants $C$ and $N_0$ such that for all $N \ge N_0$,
\begin{IEEEeqnarray*}{rCl}
 \norm{f(N)} &\le& C |g(N)| \text{\: as \:} N \to \infty.
\end{IEEEeqnarray*}
From Lemma~\ref{lem:eta_cov} it follows that
\begin{IEEEeqnarray*}{rCl}
R^m(\hat{\eta}_N,\theta) - \bar{R}(\theta) = \Oc(m(N)).
 \yesnumber \label{eq:R_m-to-R_bar-wp1}
\end{IEEEeqnarray*}
By standard continuity arguments,
%From Assumption~\ref{ass:input}, the spectral density of $\{[r(t), e(t)] \}$ is bounded below by $\delta I$.
%From Lemma 4.2 in \cite{Ljung&Wahlberg:92a}
with probability 1
\begin{IEEEeqnarray*}{rClrCl}
R^m(\hat{\eta}_N,\theta) > 0
\end{IEEEeqnarray*}
for large enough $N$. Hence, for $N$ large enough, using \eqref{eq:regression-form}  in \eqref{eq:theta_update-equation} 
\begin{IEEEeqnarray*}{rCl}
 \hat{\theta}^{k+1}_N & = &
 \theta_\nul + [R^m(\hat{\eta}_N,\theta_N^k)]^{-1}   \\
 & & \cdot \frac{1}{N}
\sum_{t = m+1}^N \varphi_t^m(\eta^n,\theta_N^k)F^\nul(q) \xi_t(\hat{\eta}_N,\hat{\theta}^k_N). \yesnumber \IEEEeqnarraynumspace \label{eq:theta-error-equation}
\end{IEEEeqnarray*}
Now, since $\{u_t\}$ is uniformly bounded and $1/F(q,\theta)$ is uniformly stable, it follows that
\begin{IEEEeqnarray*}{rCl}
 \norm{\varphi_t^m(\hat \eta_N,\theta_N^k) } \le C_1,
\end{IEEEeqnarray*}
for some $C_1 < \infty$, and furthermore, by Lemma~\ref{lem:eta_cov}, it follows that
\begin{IEEEeqnarray*}{rCl}
 F^\nul(q) \xi_t(\hat{\eta}_N,\hat{\theta}^k_N) &=& \Oc(m(N)).
\end{IEEEeqnarray*}
It thus follows that
\begin{IEEEeqnarray*}{rCl}
 \hat{\theta}^{k+1}_N - \theta_\nul  & = & \Oc(m(N)),
 \yesnumber \label{eq:theta-m(N)}
\end{IEEEeqnarray*}
for any $k \ge 0$ and 
\begin{IEEEeqnarray*}{rCl}
 \norm{ \hat{\theta}^{k+1}_N - \theta_\nul  } & \to & 0,
 \quad \mathrm{ as \: N \to \infty , \: w.p. \: 1. } 
\end{IEEEeqnarray*}

\section{Auxiliary lemmas}
\label{sec:auxiliary-lemmas}
\input{auxiliary-lemmas}

\section{Proof of Theorem~\ref{thm:asymptotic-covariance}}
\label{sec:a-not-so-long-proof}
\input{not-so-long-proof}

%% file: auxiliary-lemmas.tex
This section includes a few results needed for the proof of Theorem~\ref{thm:asymptotic-covariance} in Section~\ref{sec:a-not-so-long-proof}.

\begin{lemma}
\label{lem:straightforward}
Assume that $X(q) = \sum_{k=1}^n x_k q^{-k}$ and $Z(q) = \sum_{l=1}^n z_l q^{-l}$ are stable filters and 
%\begin{IEEEeqnarray*}{rCl}
%\sup_\omega \norm{X(\eiw) }
%,\quad \norm{Z(\eiw)}}
%\end{IEEEeqnarray*} 
let $v(t)$ be quasi-stationary.
Then,
\begin{IEEEeqnarray*}{rCl}
\norm{ \frac{1}{N}\sum_{t=m+1}^N  X(q) v(t) Z(q) v(t) }_2  & \le & \norm{X}_2 \norm{Z}_2 C
\end{IEEEeqnarray*}
for some $C < \infty $.
\end{lemma}

\begin{pf}
\begin{IEEEeqnarray*}{rl}
\Vert \frac{1}{N}\sum_{t=m+1}^N & X(q) v(t) Z(q) v(t) \bigg \Vert^2  
\\
& = \norm{ \frac{1}{N} \sum_{t=m+1}^N \sum_{k=1}^{n} x_k v_{t-k}  \sum_{l=1}^{n} z_l v_{t-l}  } ^2
\\
& = \norm{ \sum_{k=1}^{n} x_k   \sum_{l=1}^{n} z_l \frac{1}{N} \sum_{t=m+1}^N v_{t-k} v_{t-l}  } ^2
\\
& \le \sum_{k=1}^{n} |x_k|^2   \sum_{l=1}^{n} |z_l|^2 \left | \frac{1}{N} \sum_{t=m+1}^N v_{t-k} v_{t-l} \right|^2
\\
& \le \sum_{k=1}^{n} |x_k|^2   \sum_{l=1}^{n} |z_l|^2 \left | R_{vv}^N(k-l) \right|^2
\\
& \le \norm{X}_2^2 \norm{Z}_2^2 C^2 ,
\end{IEEEeqnarray*}
where the last equality is due to the quasi stationarity of $v(t)$.
\end{pf}

\begin{lemma}
\label{lem:Theorem7.3Wahlberg}
Let Assumptions~\ref{ass:true_system}, \ref{ass:input}, \ref{ass:noise}, and \ref{ass:model-order} be in force. 
Let $\Upsilon^n$ be an $m \times 2n$ deterministic matrix, with $m$ fixed. 
Then, we have that
\begin{IEEEeqnarray}{rCl}
\sqrt{N}\Upsilon^n (\hat \eta_N - \bar \eta^n) & \sim & \AsN (0, P),
\end{IEEEeqnarray}
where
\begin{IEEEeqnarray}{rCl}
P & = & \sigma_ \nul^2 \lim_{n \to \infty} \Upsilon^n [\bar R^n]^{-1} (\Upsilon^n)\transpose,
\end{IEEEeqnarray}
if the limit exists.
\end{lemma}

\begin{pf}
See \cite[Theorem 7.3]{Ljung&Wahlberg:92a}.
\end{pf}

\begin{lemma}
\label{lem:SS}
Let $\{x_n\}$ be a sequence of random variables that is asymptotically Gaussian distributed---$\{x_n\}\sim As\text{N}(0,P)$.
Let $\{M_n\}$ be a sequence of random square matrices that converge in probability to a non-singular matrix $M$, and $\{b_n\}$ be a sequence of random vectors that converges in probability to $b$.
Also, let 
\begin{equation}
y_n = M_n x_n + b_n .
\end{equation}
Then, $y_n$ converges in distribution to $\mathcal{N}(b,MPM^\top)$.
\end{lemma}

\begin{pf}
See~\cite[Lemma B.4]{soderstrom1988system}.
\end{pf}

\begin{lemma}
\label{lem:open_loop_lemma}
Let $\Sc_n$ be the subspace of $\Lc_2^2$ spanned by the rows of 
\begin{IEEEeqnarray}{rCl}
\begin{bmatrix}
-F_1 F_3 \Gamma_n   & F_2 \Gamma_n \\
 F_3 \Gamma_n & 0
\end{bmatrix} ,
\end{IEEEeqnarray}
where 
\begin{IEEEeqnarray}{rCl}
 \Gamma_n(q) &=& 
 \begin{bmatrix}
q^{-1} & \dotsc & q^{-n}
\end{bmatrix},
\\
F_i(q) &=& \sum_{k=0}^\infty f_k^i q^{-k}.
\end{IEEEeqnarray}
Suppose that $F_1,F_2$ and $F_3$ are exponentially stable, \ie for an exponentially stable $F_i$ 
\begin{IEEEeqnarray}{rCl+rCl+rCl}
|f_k^i| & \le & C \lambda^k, & \mathrm{for \,\, some \,\,} C & < & \infty, & \lambda & < & 1, \IEEEeqnarraynumspace
\end{IEEEeqnarray}
and that there is a causal exponentially stable inverse  
\begin{IEEEeqnarray}{rCl}
\tilde F_2(q) &=& \sum_{k=0}^\infty \tilde f_k^2 q^{-k} , \quad |\tilde f_k^2| < C \lambda^k .
\end{IEEEeqnarray}
Let $\gamma = [ \sum_{k=1}^\infty d_k q^{-k} \quad 0 \, ]$ be exponentially stable.
Then
\begin{IEEEeqnarray}{rCl+rCl+rCl}
\norm{\gamma - \proj[\Sc_n]{[\gamma]} }_2 & \le & C \lambda^n, & \mathrm{for \,\, some \,\,}  C & < & \infty, & \lambda & < & 1 .
\IEEEeqnarraynumspace
\end{IEEEeqnarray}

\end{lemma}

\begin{pf}
We will construct an explicit approximation to $\gamma$ that belongs to $\Sc_n$. 
Let 
\begin{IEEEeqnarray*}{rCl}
\tilde F_u \gamma &=& \begin{bmatrix}
\sum_{l=1}^\infty \beta_l z^{-l} & 0
\end{bmatrix} ,
\end{IEEEeqnarray*}
which is exponentially stable since both $\gamma$ and $\tilde F_2$ are exponentially stable. 
Take as approximation for $\gamma$
\begin{IEEEeqnarray*}{rCl}
\hat \gamma_n &=& \begin{bmatrix}
\sum_{l=1}^n \beta_l F_2(z) z^{-l} & 0
\end{bmatrix} ,
\end{IEEEeqnarray*}
which by construction belongs to $\Sc_{\Psi}$. Introduce the notation $\gamma = \begin{bmatrix} \gamma_1 &
\gamma_2
\end{bmatrix}$. Hence
\begin{IEEEeqnarray*}{rCl}
\norm{\gamma_k - \proj[\Sc_{\tilde \Psi}]{[\gamma]} }_2 &\le &  \norm{\gamma - \hat{\gamma}_n }_2
\\
&=& \norm{\gamma_1 - \sum_{l=1}^n \beta_l F_2(z) z^{-l} }_2
\\
& =  & \norm{F_2(z) \left( \tilde F_2(z) \gamma_1 - \sum_{l=1}^n \beta_l  z^{-l} \right) }_2
\\
&\le &  \norm{F_2(z)}_2 \norm{\sum_{l=n+1}^\infty \beta_l  z^{-l} }_2  \le C \lambda^n ,
\end{IEEEeqnarray*}
for some $C < \infty$ and $\lambda < 1$ since $F_2$ and $\tilde F_2 \gamma$ are exponentially stable.
\end{pf}

%% file: not-so-long-proof.tex
We start by using \eqref{eq:theta-error-equation} to write 
\begin{IEEEeqnarray*}{rCl}
 \sqrt{N}(\hat{\theta}^{k+1}_N - \theta_\nul) &=&
M_N^{-1} x_N,
\end{IEEEeqnarray*}
where
\begin{IEEEeqnarray*}{rCl}
M_N &=& R^m(\hat{\eta}_N,\theta_N^k) \\
x_N &=& \frac{1}{\sqrt{N}}
\sum_{t = m+1}^N \varphi_t^m(\hat{\eta}_N,\theta_N^k)F^\nul(q) \xi_t(\hat{\eta}_N,\hat{\theta}^k_N).
\end{IEEEeqnarray*}
From \eqref{eq:R_m-to-R_bar-wp1} and Theorem~\ref{thm:convergence}, for $k\ge 1$, we have that
\begin{IEEEeqnarray*}{rCl}
M_N \to M_{CR}, \quad \mathrm{ as \: N \to \infty , \quad w.p. \: 1.} 
\end{IEEEeqnarray*}
Assume for now (we will prove it later) that 
\begin{IEEEeqnarray*}{rCl}
x_N & \sim & \AsN(0,P).
\end{IEEEeqnarray*}
Then, using Lemma~\ref{lem:SS}, we have that
\begin{IEEEeqnarray*}{rCl}
\sqrt{N}(\hat{\theta}^{k+1}_N - \theta_\nul) & \sim & \AsN(0,M_{CR}^{-1}PM_{CR}^{-1}). \yesnumber \label{eq:Soderstrom-limit}
\end{IEEEeqnarray*}

\subsection{$x_N$}

We will now establish the asymptotic distribution and covariance of $x_N$. 
To this end, we first define
\begin{IEEEeqnarray*}{rCl}
 \Phi^m(\eta^n,\theta) & \defeq &   \frac{1}{F(q,\theta)} \begin{bmatrix}
-B(q,\eta^n) \Gamma_m \\ A(q,\eta^n) \Gamma_m
\end{bmatrix},
\\
 \Xi^m(\eta^n,\theta) & \defeq &   \frac{F^\nul(q)}{A^\nul(q)F(q,\theta)} 
 \\
 &&
 \cdot
 \begin{bmatrix}
-B^\nul(q) & A^\nul(q)
\end{bmatrix}
\begin{bmatrix}
A(q,\eta^n)- A^\nul(q) \\ B(q,\eta^n) - B^\nul(q)
\end{bmatrix} .
\end{IEEEeqnarray*}
Then we rewrite $\xi_t(\hat{\eta}_N,\theta_N^k)$ as
\begin{IEEEeqnarray*}{rCl}
\xi_t(\hat{\eta}_N,\theta_N^k) &=& -\frac{B(q,\hat{\eta}_N)}{A^\nul(q)F^(q,\theta_N^k)} (A(q,\hat{\eta}_N)-A^\nul(q))u_t
\\
&&+ \frac{A(q,\hat{\eta}_N)}{A^\nul(q)F(q,\theta_N^k)} (B(q,\hat{\eta}_N)-B^\nul(q))u_t
\\
&=&
-\frac{B^\nul(q)}{A^\nul(q)F(q,\theta_N^k)} (A(q,\hat \eta_N)-A^\nul(q))u_t
\\
&&+ \frac{A^\nul(q)}{A^\nul(q)F(q,\theta_N^k)} (B(q,\hat \eta_N)-B^\nul(q))u_t
\\
& = & \frac{1}{F^\nul(q)} \Xi^m(\hat \eta_N,\theta_N^k)u_t .
\end{IEEEeqnarray*}
We can thus express $x_N$ as
\begin{IEEEeqnarray*}{rCl}
x_N &=& \frac{1}{\sqrt{N}}
\sum_{t = m+1}^N  \Phi^m(\hat \eta_N,\theta_N^k) u_t \Xi^m(\hat \eta_N,\theta_N^k)u_t .
\end{IEEEeqnarray*}
We will in the remainder of the proof need some properties regarding the filters $\Phi^m$ and $\Xi^m$ that are easily established using Lemma~\ref{lem:eta_cov}:
\begin{IEEEeqnarray}{rCl}
\norm{ \Xi^m(\hat \eta_N,\theta_N^k)  } & = & \Oc(m(N)) \label{eq:norm-eq-1} \\
\norm{  \Phi^m(\hat \eta_N,\theta_N^k)  - \Phi^m(\hat \eta_N,\theta^\nul) } & = &  \Oc(m(N))    \label{eq:norm-eq-2}  \\
\norm{  \Phi^m(\hat \eta_N,\theta^\nul) - \Phi^m(\eta^\nul,\theta^\nul)   } & = &  \Oc(m(N))    \label{eq:norm-eq-3}  \\
\norm{  \Xi^m(\hat \eta_N,\theta_N^k)   - \Xi^m(\hat \eta_N,\theta^\nul)  } & = &  \Oc(m^2(N))  \label{eq:norm-eq-4}  \\
\norm{  \Phi^m(\eta^\nul,\theta^\nul) } & = & \Oc(1)   \label{eq:norm-eq-5}
\end{IEEEeqnarray}
For future reference, we will establish the limit of $\sqrt{N} m^2(N)$.
The dominating term in $m(N)$ is $n(N)\sqrt{\log N/N}$ and terms with $d(N)$ will be neglected. 
For $N$ large enough, we have   
\begin{IEEEeqnarray*}{rCl}
 \lim_{N \to \infty} \sqrt{N} m^2(N) & = &  \lim_{N \to \infty}  \sqrt{N} n(N)^2 \frac{\log N}{N} 
\\
&=& 
 \lim_{N \to \infty} 
\left ( \frac{n(N)^{4+\delta}}{N} \right)^{ \frac{2}{4+\delta}  }  \frac{\log N}{N^{ \frac{\delta}{4+\delta} } }  = 0 ,
\end{IEEEeqnarray*}
where the first term goes to zero by Assumption~\ref{ass:model-order}.

Using Lemma~\ref{lem:straightforward} and Lemma~\ref{lem:SS} with \eqref{eq:norm-eq-1} and \eqref{eq:norm-eq-2}, it follows that difference between $x_N$ and 
\begin{IEEEeqnarray*}{rCl}
\frac{1}{\sqrt{N}}
\sum_{t = m+1}^N  \Phi^m(\hat \eta_N,\theta_\nul) u_t \Xi^m(\hat \eta_N,\theta_N^k)u_t
\yesnumber \label{eq:straightforward-1}
\end{IEEEeqnarray*}
tend to zero as $N \to \infty$ w.p.1, and therefore they have the same asymptotic distribution and the same asymptotic covariance.
We will analyze \eqref{eq:straightforward-1} instead. 
Similarly, using Lemma~\ref{lem:straightforward} and Lemma~\ref{lem:SS} with \eqref{eq:norm-eq-1} and \eqref{eq:norm-eq-3}, it follows that difference between \eqref{eq:straightforward-1} and
\begin{IEEEeqnarray*}{rCl}
\frac{1}{\sqrt{N}}
\sum_{t = m+1}^N  \Phi^m(\eta^\nul,\theta_\nul) u_t \Xi^m(\hat \eta_N,\theta_N^k)u_t
\yesnumber \label{eq:straightforward-2}
\end{IEEEeqnarray*}
tend to zero as $N \to \infty$ w.p.1, and we will analyze \eqref{eq:straightforward-2} instead. 
Similarly, using Lemma~\ref{lem:straightforward} and Lemma~\ref{lem:SS} with \eqref{eq:norm-eq-4} and \eqref{eq:norm-eq-5}, the difference between \eqref{eq:straightforward-2} and
\begin{IEEEeqnarray*}{rCl}
\frac{1}{\sqrt{N}}
\sum_{t = m+1}^N  \Phi^m(\eta^\nul,\theta_\nul) u_t \Xi^m(\hat \eta_N,\theta^\nul)u_t
\yesnumber \label{eq:straightforward-3}
\end{IEEEeqnarray*}
tend to zero as $N \to \infty$ w.p.1, and we will analyze \eqref{eq:straightforward-3} instead. 

We rewrite $\Xi^m(\hat \eta_N,\theta^\nul)u_t$ as
\begin{IEEEeqnarray*}{rCl}
\Xi^m(\hat \eta_N,\theta^\nul)u_t &=& \frac{1}{A^\nul(q)}
\begin{bmatrix}
-B^\nul(q)u_t \Gamma_n \\ A^\nul(q)u_t \Gamma_n
\end{bmatrix}\transpose
\!\! (\hat{\eta}_N - \bar{\eta}^n)
\\
&=&
\frac{1}{A^\nul(q)} \varphi^n_t(\eta_\nul,\theta^\nul)\transpose
(\hat{\eta}_N - \bar{\eta}^n) .
\yesnumber
\end{IEEEeqnarray*}
Thus, we have shown that $x_N$ has the same distribution and covariance as 
\begin{IEEEeqnarray}{rCl}
T_N &\defeq& Z^{n} \sqrt{N} (\hat{\eta}_N - \bar{\eta}^n) , \IEEEeqnarraynumspace \label{eq:T_N-def}
\end{IEEEeqnarray}
where
\begin{IEEEeqnarray*}{rCl}
\label{eq:Z-def}
Z^{n} &=&  \sum_{t=m+1}^N \varphi^m_t(\eta_\nul,\theta_\nul) \frac{F^\nul(q)}{A^\nul(q)} \varphi^n_t(\eta_\nul,\theta_\nul)\transpose , \yesnumber
\end{IEEEeqnarray*}
and we will analyze $T_N$ instead.

\subsection{Asymptotic covariance of $T_N$}
\label{sec:T-as-var}
Using Lemma~\ref{lem:Theorem7.3Wahlberg}, we have that 
\begin{IEEEeqnarray*}{rCl}
T_N & \sim & \AsN(0,Q),
\end{IEEEeqnarray*}
where
\begin{IEEEeqnarray*}{rCl}
Q = \sigma_\nul^2 \lim_{n \to \infty} Z^n[\bar R^n]^{-1} (Z^n)\transpose, \yesnumber \label{eq:Q-def}
\end{IEEEeqnarray*}
provided the right hand side limit exists. 
This will be shown next. 
We start by analyzing $\bar{R}^{n}$. 
\begin{IEEEeqnarray*}{rCl}
\bar{R}^{n} &=&
\expect{ \varphi^n_t(\varphi^n_t )\transpose }
\\
&=&
\inp{ \Psi, \Psi } ,
\yesnumber \IEEEeqnarraynumspace \label{eq:Psi_Psi}
\end{IEEEeqnarray*}
where
\begin{IEEEeqnarray*}{rCl}
\inp{ f, g }  &\coloneqq& 
\wint{f(e^{j\omega})g(e^{j\omega})^*}
,
\end{IEEEeqnarray*}
and with $\Psi$ given by
\begin{IEEEeqnarray*}{rCl}
\Psi &=& \begin{bmatrix}
-G^\nul \Gamma_n  & H^\nul \Gamma_n \\ \Gamma_n & 0_{n \times 1}
\end{bmatrix} U_\nul
\end{IEEEeqnarray*}
and $U_\nul$ is a spectral factor of the the covariance matrix of the input $u_t$ and the noise $e_t$, given by 
\begin{IEEEeqnarray*}{rCl}
U_\nul &=& \begin{bmatrix}
F_u  &   0 \\
0 & \sigma_\nul
\end{bmatrix} .
\end{IEEEeqnarray*}
For \eqref{eq:Z-def}, we have that
\begin{IEEEeqnarray*}{rCl}
\label{eq:Z-calc}
Z^{n} &=& \expect{ \varphi^m_t(\eta_\nul,\theta_\nul) \frac{F^\nul(q)}{A^\nul(q)} \varphi^n_t(\eta_\nul,\theta_\nul)\transpose }
\\
&=&
\expect{
\begin{bmatrix}
- \frac{B^\nul}{F^\nul}\Gamma_m u_t \\
  \frac{A^\nul}{F^\nul}\Gamma_m u_t
\end{bmatrix}
\begin{bmatrix}
-G^\nul \Gamma_n u_t  \\ \Gamma_n u_t
\end{bmatrix}\transpose
}
\\
&=&
\inp{
\begin{bmatrix}
- \frac{G^\nul}{F^\nul H^\nul}\Gamma_m & 0_{n \times 1}  \\
  \frac{1}{F^\nul H^\nul}\Gamma_m & 0_{n \times 1}
\end{bmatrix}
F_u
,
\begin{bmatrix}
-G^\nul  \Gamma_n  & 0_{n \times 1} \\ 
 \Gamma_n & 0_{n \times 1}
\end{bmatrix}
F_u
}
\\
&=& \inp{\gamma,  \Psi} ,
\yesnumber \IEEEeqnarraynumspace \label{eq:gamma_psi}
\end{IEEEeqnarray*}
with
\begin{IEEEeqnarray*}{rCl}
\gamma &=& 
\begin{bmatrix}
- \frac{G^\nul}{F^\nul H^\nul}\Gamma_m & 0_{m \times 1}  \\
  \frac{1}{F^\nul H^\nul}\Gamma_m & 0_{m \times 1}
\end{bmatrix} F_u,
\end{IEEEeqnarray*}
where the last equality is due to the fact that the added column in the right argument of the inner product is multiplied by the zero column in $\gamma$ when the inner product is taken. 
Hence, we can write the asymptotic covariance matrix of $T_N$ as 
\begin{IEEEeqnarray*}{rCl}
\lim_{N \to  \infty} \expect{T_NT_N \transpose}
&=& \sigma_\nul^2 
\inp{\gamma, \Psi } \inp{ \Psi , \Psi }^{-1} \inp{\Psi ,  \gamma } 
\\
&=& \sigma_\nul^2 
\inp{\proj[\Sc_\Psi]{[\gamma]} , \proj[\Sc_\Psi]{[\gamma]} } ,
\yesnumber \IEEEeqnarraynumspace
\label{eq:Psi_projection}
\end{IEEEeqnarray*}
where $\Sc_\Psi$ is the subspace in $\Lc_2^{1 \times 2}$ spanned by the rows of $\Psi$.
Lemma~\ref{lem:open_loop_lemma} gives that, as $n \to \infty$, $S_{ \gamma } \subseteq \Sc_\Psi$ and
\begin{IEEEeqnarray*}{rCl}
\lim_{N \to  \infty} \expect{T_NT_N \transpose}
&=&  \sigma_\nul^2 
\inp{\gamma, \gamma }  = \sigma_\nul^2 M_{CR} .
\label{eq:T-limit}
\end{IEEEeqnarray*}

\subsection{Summing up}

Consider $T_N$ defined in \eqref{eq:T_N-def}.
As observed in Section~\ref{sec:T-as-var}, it follows from Lemma~\ref{lem:Theorem7.3Wahlberg} that
\begin{IEEEeqnarray}{rCl}
\label{eq:T-normality}
T_N &\sim & \AsN(0, \sigma_\nul^2 M_{CR}).
\end{IEEEeqnarray}
The asymptotic normality of $\sqrt{N} (\hat{\theta}_N  - \hat{\theta}_\nul)$ follows from~\eqref{eq:Soderstrom-limit}
and~\eqref{eq:T-normality}, together with that $\sqrt{N} (\hat{\theta}_N  - \hat{\theta}_\nul)$ has the same asymptotic distribution as $T_N$.
From ~\eqref{eq:Soderstrom-limit} and \eqref{eq:T-normality}, it now follows that 
\begin{IEEEeqnarray*}{rCl}
\sqrt{N}( \hat{\theta}_N^k - \theta_ \nul) & \sim & \AsN(0,\sigma_\nul^2 M_{CR}^{-1}).
\IEEEeqnarraynumspace
\yesnumber 
\end{IEEEeqnarray*}